\documentclass[12pt]{article}
\usepackage{epsfig}
\title{Chiral effective theory of strong interactions.}
\vspace{2cm}
\author{B.L.Ioffe\\
\\
 Institute of Theoretical and Experimental Physics\\
B.Cheremushkinskaya 25, 117218 Moscow,Russia}
\date{}
\begin{document}

\maketitle

\newcommand{\be}{\begin{equation}}
\newcommand{\ee}{\end{equation}}

\def\la{\mathrel{\mathpalette\fun <}}
\def\ga{\mathrel{\mathpalette\fun >}}
\def\fun#1#2{\lower3.6pt\vbox{\baselineskip0pt\lineskip.9pt
\ialign{$\mathsurround=0pt#1\hfil##\hfil$\crcr#2\crcr\sim\crcr}}}

\vspace{1cm}

\begin{abstract}

The review of chiral effective theory (CET) is given. CET is based on
quantum chromodynamics and describes the  processes of strong interaction at
low energies. It is proved, that CET comes as a consequence of the
spontaneous violation of chiral symmetry in QCD -- the appearance of chiral
symmetry violating vacuum condensates. The Goldstone theorem for the case of
QCD is proved and the existence of the octet of massless Goldstone bosons
($\pi,K,\eta$) is demonstrated  in the limit of massless
$u,d,s$  quarks (or the triplet of massless pions in the limit $m_u,m_d \to
0$). It is shown, that the same phenomenon -- the appearance of quark
condensate in QCD -- which causes the Goldstone bosons, results in
appearance of violating chiral symmetry massive baryons. The general form of
CET Lagrangian is derived. Few examples of higher order corrections to tree
diagrams in CET are given. The Wess-Zumino term (of order $p^4$ term in CET
Lagrangian)  is presented. Low energy sum rules are presented. QCD and CET
at finite temperature are discussed. In the framework of CET the $T^2$
correction to quark condensate in QCD at finite temperature $T$  is
calculated and the results of higher order temperature corrections are
demonstrated. These results indicate on phase  transition in QCD at
$T\simeq 150-200$ MeV. The mixing of current correlators in order $T^2$ is
proved.

\vspace{5mm}


\vspace{5mm}

\centerline{Review submitted for publication to "Uspekhi Fiz. Nayk"}

\end{abstract}

\newpage

\centerline{\bf Contents}

\vspace{10mm}

1. Introduction.

2. The masses of light quarks.

3. Spontaneous violation of chiral symmetry. Quark condensate.

4. Goldstone theorem.

5. Nucleon mass and quark condensate.

6. Chiral effective theory at low energies.

7. Low energy sum rules in CET.

8. QCD and CET at finite temperature.

9. Conclusion.

10. References.

\vspace{10mm}

e-mail: ioffe@vitep5.itep.ru

\newpage

\section{Introduction}

It is generally accepted -- and this fact is undisputable now -- that the
true theory of strong interactions is quantum chromodynamics (QCD),
nonabelian gauge theory of interacting quarks and gluons. QCD possesses a
striking property of asymptotic freedom: the coupling constant
$\alpha_s(Q^2)$ decreases logarithmically as a function of momentum transfer
square $Q^2$  at large $Q^2$: $\alpha_s(Q^2)\sim 1/ln~Q^2$  at $Q^2 \to
\infty$ (or, what is equivalent, $\alpha_s$ decreases at small distances,
$\alpha_s(r)\sim 1/lnr$). This property of QCD allows one to perform
reliable theoretical calculation of the processes, proceeding at high
momentum transfers (at small distances) by using perturbation theory.
However, the same property of the theory involves the increase (in the
framework of perturbative theory an unlimited ones) of the running coupling
constant in QCD at small momentum transfer, i.e. at large distances.
Physically, such growth is natural and, even more, it is needed,  otherwise
the  theory would not be a theory of \underline{strong}  interactions. QCD
possesses also the other remarkable property, the property of confinement:
quarks and gluons cannot leave the region of their strong interaction  and
cannot be observed as real physical objects. Physical objects, observed
experimentally at large distances, are hadrons--mesons and baryons. These
two  circumstances -- the growth of coupling constant and the phenomenon of
confinement, do not make it possible as a rule, to predict theoretically in
QCD the processes at low energies and the properties of physical hadrons.
(Some exceptions from this rule are low energy theorems, proved in QCD,
and,  especially the powerful QCD sum rule method, which, although, is
starting from small distances, but allows one in many cases to go to rather
large ones and to describe the properties  of hadrons. The other exceptions
are the numerical calculations on lattices.)

However, it became possible to construct the QCD based effective theory,
which describes the processes of strong interaction at low energies. The
small parameters in the theory are the momenta of interacting particles,
more exactly, their ratios to the characteristic hadronic mass scale $M\sim
0.5 - 1$ GeV, $p_i/M \ll 1$.  The theory is constructed as a series in the
powers of $p_i/M$  and is an effective theory, i.e., when going to the next
order terms in $p_i/M$, in the Lagrangian one must take into account,
additional terms, characterized by new parameters. The appearance of
effective theory is connected with one more specific property of QCD -- the
spontaneous breaking of chiral symmetry. The masses of light $u,d,s$ quarks
which enter the QCD Lagrangian, especially the masses of $u$  and $d$
quarks, from which the usual (nonstrange)  hadrons are built, are very small
as composed with characteristic mass scale $m_u, m_d < 10$ MeV. Since in QCD
the quark interaction proceeds  through the exchange of vector gluonic
field, then, if light quark masses are neglected, QCD Lagrangian (its light
quark part) is chirally symmetric, i.e. not only vector, but also axial
currents are conserved and the left and right chirality quark fields are not
interacting with one another. This chiral symmetry is not realized in the
spectrum of hadrons and their low energy interactions. Indeed, in chirally
symmetrical theory the fermion states must be either the massless or
degenerate in parity. It is evident, that the baryons (particularly, the
nucleon) do not possess such properties. This means, that the chiral
symmetry of QCD Lagrangian is not realized on the spectrum of physical
states and is spontaneously broken. According to Goldstone theorem
spontaneous breaking of symmetry leads to appearance of massless particles
in the spectrum of physical states -- the Goldstone bosons. In QCD Goldstone
bosons may be indentified with the triplet of $\pi$-mesons in the limit
$m_u,m_d \to 0,$ $m_s\not= 0$ ($SU(2)$--symmetry)  and the octet of
pseudoscalar mesons $(\pi, K,\eta)$  in the limit $m_u,m_d,m_s \to 0$
($SU(3)$-symmetry). The local $SU(2)_V \times SU(2)_A$  symmetry (here $V$
and $A$ mean  vector and axial currents, $u$ and $d$ quarks are considered
as massless)  or $SU(3)_V\times SU(3)_A$ symmetry (if $m_s$ is also
neglected) of hadronic strong interaction and the existence of massless
Goldstone bosons allows one to construct the effective chiral theory of
Goldstone bosons and their interactions with baryons, which is valid at
small particle momenta.

In the initial version, before QCD, this approach was called the theory of
partial conservation of axial current (PCAC). The Lagrangian of the theory
represented the nonlinear interaction of pions with themselves and with
nucleons and corresponded to the first term in the  expansion in powers of
momenta in modern chiral effective theory. (The review of the PCAC theory at
this stage was done in [1]). When QCD  had been created, it was proved, that
the appearance of Goldstone bosons is a consequence of spontaneous breaking
of chiral symmetry in QCD vacuum and it is tightly connected with the
existence of vacuum condensates, violating the chiral symmetry. It had been
also established, that baryon masses are expressed through the same vacuum
condensates. Therefore, basing on QCD, the mutual interconnection of all set
of phenomena under  consideration was found. It was possible to formulate
the chiral effective theory (CET)  of hadrons as a succesive expansion in
powers of particle momenta and quark (or Goldstone bosons) masses not only
in tree approximation, as in PCAC, but also with account of loops. (CET is
often called chiral perturbation theory -- ChPT.)

In this review  the foundations, basic ideas and concepts of CET are
considered as well as their connection with QCD. The main attention is paid
to the general properties of pion interactions. For pion-nucleon interaction
only the general form of Lagrangian is presented. The physical effects are
considered as illustrative examples in nonsystematical way. In fact in CET a
lot of such effects was calculated (particularly, for meson--baryon
interactions, meson and baryon formfactors etc.)  They are very interesting
for specialists, but their inclusion into review would increase its space
drastically. The comparison of the theory with experiment almost will not be
discussed. Such discussion could be a subject of a separate review.

\section{The masses of the light quarks}

In what follows $u, d, s$ quarks will be called as "light quarks" and all
other quarks as "heavy quarks". The reason is that the masses  of the light
quarks are small compared with the characteristic mass of strong
interaction $M \sim 0.5 - 1.0$ GeV or $m_{\rho}$. This statement is a
consequence of the whole set of facts confirming that the symmetry of strong
interaction is $SU(3)_L \times SU(3)_R \times U(1)$. Here the group
generators are the charges corresponding to the left $(V-A)$ and right
$(V+A)$ light quark chiral currents and $U(1)$ corresponds to the baryonic
charge current. The experiment shows that the accuracy of $SU(3)_L \times
SU(3)_R$ symmetry is of the same order as the accuracy of the $SU(3)$
symmetry: the small parameter characterizing the chiral symmetry violation
in strong interactions is generally of order $\sim 1/5 - 1/10$.

The approximate validity of the chiral symmetry means that not only
divergences of the vector currents $\partial_{\mu} j^q_{\mu}$ are zero or
small, but also of the axial ones $\partial_{\mu} j^q_{\mu 5}$. (Here $q =
u,d,s$. This statement refers to nonsinglet in flavor axial currents. The
 divergence of singlet axial current is determined by the anomaly and is
nonzero even for massless quarks -- the discussion of this problem is
outside the scope of this review).  The divergences of nonsinglet axial
currents in QCD are proportional to quark masses.  Therefore the existence
of the chiral symmetry can be understood if the quark masses are small
\cite{1,2}.  However, the baryon masses are by no means small:  the chiral
symmetry is not realized on the hadronic mass spectrum in a trivial way by
vanishing of all the fermion masses. This means that the chiral symmetry is
broken spontaneously by the physical states spectrum.  According to the
Goldstone theorem such a symmetry breaking results in appearance of massless
particles -- Goldstone bosons. In the case considered these Goldstone bosons
must belong to a pseudoscalar octet. They are massless if quark masses are
put to zero. The nonvanishing quark masses realize the explicit violation of
the chiral symmetry and provide the masses of the pseudoscalar meson octet.
For this reason the pseudoscalar meson octet (often called the octet of the
Goldstone bosons) plays a special role in QCD.

Heavy quarks are decoupled in the low energy domain (this statement is
called the Appelquist-Carazzone theorem) \cite{6}. We ignore them in this
Chapter where QCD at low energies is considered.  The QCD Hamiltonian can be
split into two pieces


\be
H = H_0 + H_1
\label{2.1}
\ee
where
\be
H_1 = \int d^3 x (m_u \bar{u} u + m_d \bar{d} d + m_s \bar{s} s)
\label{2.2}
\ee
Evidently, because of vector gluon-quark interaction
the first term in Hamiltonian -- $H_0$ is $SU(3)_L
\times SU(3)_R$  invariant and the only source of $SU(3)_L\times SU(3)_R$
violation is $H_1$.
The quark masses $m_q, q = u,d,s$ in (\ref{2.2}) are not renormalization
invariant:  they are scale dependent. It is possible to write


\be
m_q (M) = Z_q(M/\mu) m_q(\mu)
\label{2.3}
\ee
where $M$ characterizes the scale, $\mu$ is some fixed normalization point
and $Z_q(M/\mu)$ are renormalization factors. If the light quark masses are
small and can be neglected, the renormalization factors are
flavor-independent and the ratios

\be
\frac{m_{q1} (M)}{m_{q2}(M)} = \frac{m_{q1} (\mu)}{m_{q2}(\mu)}
\label{2.4}
\ee
are scale-independent and have definite physical meaning. (This relation
takes place if $M$ is higher than the Goldstone mass $m_k$ : its validity in
the domain $M \sim m_k$ may lead to some doubts).

In order to find the ratios $m_u/m_d$ and $m_s/m_d$ consider the axial
currents

$$
j^-_{\mu 5} = \bar{d} \gamma_{\mu} \gamma_5 u
$$
\be
j^3_{\mu 5} = [\bar{u} \gamma_{\mu} \gamma_5 u - \bar{d} \gamma_{\mu}
\gamma_5 d] /\sqrt{2}
\label{2.5}
\ee
\be
j^{s-}_{\mu 5} = \bar{s} \gamma_{\mu} \gamma_5 u, ~~~ j^{s 0}_{\mu 5} =
\bar{s}\gamma_{\mu} \gamma_5 d
\label{2.6}
\ee
and their matrix elements between vacuum and $\pi$ or $K$ meson states.

$$
\langle 0 \mid j^-_{\mu 5} \mid \pi^+ \rangle = if_{\pi^+} p_{\mu}
$$
$$
\langle 0 \mid j^{3}_{\mu 5} \mid \pi^0 \rangle = if_{\pi^0} p_{\mu}
$$
$$
\langle 0 \mid j^{s-}_{\mu 5} \mid K^+ \rangle = if_{K^+} p_{\mu}
$$
\be
\langle \mid j^{s 0}_{\mu 5} \mid K^0\rangle  = if_{K^0} p_{\mu}
\label{2.7}
\ee
where $p_{\mu}$ are $\pi$ or $K$ momenta. In the limit of strict $SU(3)$
symmetry all constants in the rhs of (\ref{2.7})  are equal: $f_{\pi^+
}=f_{\pi^0}=f_{K^+}=f_{K^0}$, $SU(2)$ -- isotopical symmetry results to
equalities $f_{\pi^+}=f_{\pi^0},~f_{K^+}=f_{K^0}$. The constants
$f_{\pi}\equiv f_{\pi}$  and $f_{K^+}\equiv f_K$ have the meaning of
coupling constants in the decays $\pi^+\to \mu^+\nu$  and $K^+\to \mu^+\nu$.
Experimentally they are equal to $f_{\pi}=131~MeV$, $f_K=160~MeV$. The ratio
$f_K/f_{\pi}=1.22$  characterizes the accuracy of $SU(3)$ symmetry. Multiply
(\ref{2.7}) by $p_{\mu}$. Using the equality for the divergence of axial
current following from QCD Lagrangian

\be
\partial_{\mu}[\bar{q}_1(x)\gamma_{\mu}\gamma_5 q_2(x)] = i(m_{q_1}+m_{q_2})
\bar{q}_1(x)\gamma_5q_2(x),
\label{2.8}
\ee
we get

$$i(m_u+m_d)\langle 0 \mid \bar{d}\gamma_5 u\mid \pi^+\rangle =
f_{\pi^+} m^2_{\pi^+} $$
$$(i/\sqrt{2})[(m_u+m_d) \langle 0\mid\bar{u}\gamma_5 u
-\bar{d}\gamma_5 d \mid \pi^0\rangle + (m_u-m_d) \langle
 0\mid\bar{u}\gamma_5 u +\bar{d}\gamma_5 d \mid
\pi^0\rangle]=f_{\pi^0}m^2_{\pi^0}$$
$$i(m_s +m_u)\langle 0\mid\bar{s}\gamma_5 u \mid K^+ \rangle =f_{K^+}
m^2_{K^+}$$
\be
i(m_s +m_d)\langle 0\mid\bar{s}\gamma_5 d \mid K^0 \rangle =f_{K^0}
m^2_{K^0}
\label{2.9}
\ee
Neglect electromagnetic (and weak) interaction and assume that isotopic
invariance may be used for the matrix elements in the lhs of (\ref{2.9}).
Then

$$\langle 0\mid\bar{u}\gamma_5 u +\bar{d}\gamma_5 d \mid
\pi^0\rangle =0,$$
\be
\langle 0\mid\bar{d}\gamma_5 u \mid \pi^+\rangle = \frac{1}{\sqrt{2}}
\langle 0\mid\bar{u}\gamma_5 u -\bar{d}\gamma_5 d \mid
\pi^0\rangle
\label{2.10}
\ee
and, as follows from (\ref{2.9}), $\pi^{\pm}$ and $\pi^0$   masses are
equal in this approximation even when $m_u\not=m_d$. Hence the
experimentally observed mass difference $\Delta
m_{\pi}=m_{\pi^+}-m_{\pi^0}=4.6~MeV$  is caused by the electromagnetic
interaction only.  The sign of the $K$-meson mass difference $\Delta
m_K=m_{K^+}-m_{K^0}=-4.0~MeV$ is opposite to that of the pion ones.
The electromagnetic kaon and pion mass differences in QCD or in the quark
model are determined by the same diagrams, and must, at least, to be of the
same sign.  This means, in accord with (\ref{2.9}), that $m_d > m_u$.

Assuming the $SU(3)$ invariance of matrix elements in (\ref{2.9}) and using
simple algebra, it is easy to get from  (\ref{2.9}) and (\ref{2.10})

$$\frac{m_u}{m_d} = \frac{\bar{m}^2_{\pi}-(\bar{m}^2_{K^0}-\bar{m}^2_{K^+})}
{\bar{m}^2_{\pi}+({\bar{m}^2_{K^0}}-\bar{m}^2_{K^+})}$$
\be
\frac{m_s}{m_d} = \frac{\bar{m}^2_{K^0}+\bar{m}^2_{K^+}-\bar{m}^2_{\pi}}
{\bar{m}^2_{K^0}-{\bar{m}^2_{K^+}}+\bar{m}^2_{\pi}}
\label{2.11}
\ee
The bars in (\ref{2.11})  mean, that the pion and kaon masses here are not
the physical ones, but the masses in the limit, when the electromagnetic
interaction is switched off. In order to relate
$\bar{m}^2_{\pi},~\bar{m}^2_K$ to physical masses, let us use again
the $SU(3)$ symmetry. In the $SU(3)$  symmetry the photon is $U$-scalar and
$\pi^+$  and $K^+$  belong to the $U$  doublet.  Therefore, the
electromagnetic corrections to $m^2_{\pi^+}$ and $m^2_{K^+}$ are equal

\be
(\delta m^2_{\pi^+})_{el} = (\delta m^2_{K^+})_{el}
\label{2.12}
\ee
It can be shown also, that in the limit $m^2_{\pi},~m^2_K\to 0$, the
electromagnetic corrections to the $\pi^0$ and $K^0$ masses tend to zero,

\be
(\delta m^2_{\pi^0})_{el}=(\delta m^2_{K^0})_{el} = 0
\label{2.13}
\ee
Eq.'s (\ref{2.12}), (\ref{2.13}) may be rewritten in the form of the Dashen
relation \cite{3}

\be
(m^2_{\pi^+} -m^2_{\pi^0})_{el} = (m^2_{K^+}-m^2_{K^0})_{el}
\label{2.14}
\ee
From (\ref{2.13}), (\ref{2.14}) we have

$$\bar{m}^2_{\pi} = m^2_{\pi^0}$$
\be
\bar{m}^2_{K^+}-\bar{m}^2_{K^0} = m^2_{K^+} - m^2_{K^0} -
(m^2_{\pi^+}-m^2_{\pi^0})
\label{2.15}
\ee
The substitution of (15) into (\ref{2.11}) leads to:

$$\frac{m_u}{m_d} = \frac{2m^2_{\pi^0}-m^2_{\pi^+} -(m^2_{K^0}-m^2_{K^+})}
{m^2_{K^0}-m^2_{K^+}+m^2_{\pi^+}} $$
\be
\frac{m_s}{m_d} = \frac{m^2_{K^0}+m^2_{K^+} -m^2_{\pi^0}}
{m^2_{K^0}-m^2_{K^+}+m^2_{\pi^+}}
\label{2.16}
\ee
Numerically, this gives [6,7]

\be
\frac{m_u}{m_d} = 0.56,~~~~~~\frac{m_s}{m_d}=20.1
\label{2.17}
\ee
A strong violation of isotopic invariance, as well as large difference
between $u,d$ and $s$-quark masses, i.e. the violation of $SU(3)$ flavor
symmetry, is evident from (\ref{2.17}). (A more detailed analysis shows,
that the results (\ref{2.17}) only slightly depend on the assumption
of the $SU(3)$ symmetry of the corresponding matrix elements used in their
derivation.)This seems to be in contradiction with the well established
isospin symmetry of strong interaction, as well as with the approximate
$SU(3)$ symmetry.The resolution of this puzzle is that the quark masses are
small: the parameter characterizing isospin violation is $(m_d - m_u)/M$ and
the parameter characterizing the $SU(3)$ symmetry violation is $m_s/M$.

Estimation of the absolute value of the quark masses can be obtained in the
following way. Suppose that the hadrons which contain strange quarks and
which belong to a given unitary multiplet are heavier only because of the
strange quark mass. Then from consideration of mass splittings in the baryon
octet one can found that $m_s \approx 150 MeV$ at a scale of about 1 GeV.
From
(\ref{2.17}) it then follows

\be
m_u = 4.2 MeV,~~~ m_d = 7.5 MeV, ~~~ m_s = 150 MeV
\label{2.18}
\ee
at 1 GeV. Taking these values, one may expect that isospin violation could
be of order $(m_d - m_u)/M \sim 10^{-2}$, i.e. of the same order as arising
from electromagnetic interaction. The order of the $SU(3)$ symmetry
violation is $\sim m_s/M \sim 1/5$.  The large $m_s/m_d$ ratio explains the
large mass splitting in the pseudoscalar meson octet. Assuming SU(3)
symmetry of the matrix elements in (\ref{2.9}), we have

\be
\frac{m^2_{K^+}}{m^2_{\pi^+}}= \frac{m_s + m_u}{m_d + m_u} = 13
\label{2.19}
\ee
in agreement with the experiment.

An important consequence of eq.'s (\ref{2.9}) is that in QCD the mass
squares of the pseudoscalar meson octet $m^2_{\pi}, m^2_K, m^2_{\eta}$ are
proportional to the quark masses and vanish when $m_q$ tend to zero: in
this limit the octet of pseudoscalar mesons becomes massless.

\section{Spontaneous violation of chiral symmetry}

\vspace{3mm}

\hspace{4.5cm} {\bf \large Quark condensate}

\vspace{5mm}
As has been already mentioned, the large baryon masses indicate that chiral
symmetry in QCD is broken spontaneously. Indeed, let us consider any process
with participation of a polarized baryon, e.g., any hadron-proton scattering
on a longitudinally polarized proton at energies of order 1 GeV. We can
treat the initial state of a polarized proton as a state with some fixed
quark helicities. Due to the chiral symmetry the helicities are conserved in
the course of collision. Therefore, we could expect that proton longitudinal
polarization will not change in the collision. However, it is
well known -- and this is a direct consequence of the Dirac equation for
proton -- that the proton mass results in proton helicity flip with not a
small probability in such a process. This simple fact -- the existence of
the proton mass -- clearly  demonstrates the violation of the chiral
symmetry in strong interactions at low energies. \footnote{In principle, the
chiral symmetry in baryonic states could be realized in a way that all
baryonic states would be degenerated in parity with a splitting of order of
$m_u + m_d$. This is evidently not the case.}

In all  known examples of the field theories the spontaneous violation of
global symmetry manifests itself in the modification of the properties of
the ground state -- the vacuum. Let us show that such phenomenon takes place
also in QCD.

Consider the matrix element

\be
i q_{\mu} (m_u + m_d)~ \int~ d^4 x e^{iqx} \langle 0 \mid T \{ j^-_{\mu 5}
(x), ~\bar{u} (0) \gamma_5  d (0) \}\mid 0 \rangle
\label{2.20}
\ee
in the limit of massless $u$ and $d$ quarks (except for the overall factor
$m_u + m_d$ ). Put $q_{\mu}$ inside the integral, integrate in parts and use
the conservation of the axial current. Then only the term with the equal
time commutator will remain

$$
- (m_u + m_d)~ \int d^4 x e^{iqx} \langle 0 \mid \delta (x_0)
 [~j^-_{0 5} (x), ~~ \bar{u} (0) \gamma_5 d (0)~ ] \mid 0 \rangle =
$$
\be
= (m_u + m_d) \langle 0 \mid \bar{u} u + \bar{d} d \mid 0 \rangle
\label{2.21}
\ee
Let us go now to the limit $q_{\mu} \to 0$ in (\ref{2.20}) and perform the
 summation over all intermediate states.  The nonvanishing contribution
 comes only from one pion intermediate state, since in this approximation
 the pion should be considered as massless. This contribution is equal to

\be
q_{\mu} \langle 0\mid j^-_{\mu 5} \mid \pi^+ \rangle
 \frac{-1}{q^2}~ \langle \pi^+ \mid (m_u + m_d) \bar{u} \gamma_5 d \mid 0
 \rangle = -f^2_{\pi} m^2_{\pi},
\label{2.22}
\ee
where (\ref{2.7}) and
 (\ref{2.9}) where substituted when going to the right  hand side. Putting
(22) in the left hand side of (21) we get

\be
\langle 0 \mid \bar{q} q \mid 0 \rangle = -\frac{1}{2} ~ \frac{m^2_{\pi}
f^2_{\pi}}{m_u + m_d},
\label{2.23}
\ee
where $q = u$ or $d$ and SU(2) invariance of QCD vacuum were used.
Eq.(\ref{2.23}) is called  Gell-Mann-Oakes-Renner relation \cite{7}. It can
be also derived in an other way. Assume the quark masses to be nonzero, but
small. Then the pion is massive and (\ref{2.20}) tends to zero in the limit
$q_{\mu} \to 0$ . However, when we put $q_{\mu}$ inside the integral, in
addition to the equal time commutator term (\ref{2.21}), the term with the
divergence of the axial current will appear. The account of this term,
saturated by one pion intermediate state, results in the same
eq.(\ref{2.23}). Numerically, with the quark mass values (\ref{2.18}) we
have

\be
\langle 0 \mid \bar{q} q \mid 0 \rangle = - (240 MeV)^3
\label{2.24}
\ee
As follows from (\ref{2.23}), the product $(m_u + m_d) \langle 0\mid \bar{q}
q \mid 0 \rangle$ is scale independent, while $\langle 0\mid \bar{q} q \mid
0 \rangle$  depends on the scale and the numerical value (\ref{2.24}) refers
to 1 GeV. The quantity $\langle 0 \mid \bar{q} q \mid 0 \rangle$, called
vacuum quark condensate can be also represented as

\be
\langle 0 \mid \bar{q} q \mid 0 \rangle = \langle 0 \mid \bar{q}_L q_R +
\bar{q}_R q _L \mid 0 \rangle
\label{2.25}
\ee
where $q_L$ and $q_R$ are left and right quark fields
$q_L = (1/2) (1 + \gamma_5)q, ~~~ q_R = (1/2)(1 - \gamma_5) q$.
It is evident from (\ref{2.24}) that quark condensate violates chiral
invariance and its numerical value (\ref{2.24}) has a characteristic
hadronic scale. The chiral invariance is violated globally, because
$\langle 0 \mid \bar{q} q \mid 0~ \rangle$ is noninvariant under global
transformations $q \to e^{i \alpha \gamma_5} q$ with a constant $\alpha$.

Surely, in perturbative QCD with massless quarks the quark condensate is
zero in any order of perturbation theory. Therefore, the nonzero and
non-small value of the quark condensate may arise only due to
nonperturbative effects. The conclusion is, that the nonperturbative field
fluctuations which violate chiral invariance of the Lagrangian, are present
and essential in QCD. Quark condensate plays a special role because its
lowest dimension, d = 3.

\section{Goldstone theorem}

In Sec.2.2 we have presented two arguments in favor that chiral
symmetry,  approximately valid in QCD because of small $u, d, s$
quark masses, is  spontaneously broken. These arguments were: the existence
of
large baryon masses and the appearance of violating chiral symmetry quark
condensate. Let us go to the limit of massless $u,d,s$ quarks and show now
that the direct consequence from each of these arguments is the appearance
of massless pseudoscalar bosons in the hadronic
spectrum.

Consider  the matrix element of the axial current $j^+_{\mu 5} = \bar{u}
\gamma_{\mu} \gamma_5 d$ between the neutron and proton states. The general
from of this matrix element is:

\be
\langle p \mid j^+_{\mu 5} \mid n \rangle = \bar{v}_p (p^{\prime}) \Biggl
[\gamma_{\mu} \gamma_5 F_1 (q^2) + q_{\mu} \gamma_5 F_2 (q^2) \Biggr ] v_n
(p),
\label{2.26}
\ee
where $p$ and $p^{\prime}$ are neutron and proton momenta, $q = p^{\prime} -
p, ~~ v_p(p^{\prime}), ~~ v_n(p)$ are proton and neutron spinors and
$F_1(q^2), ~ F_2(q^2)$ are formfactors. Multiply (\ref{2.26}) by $q_{\mu}$
and go to the limit $q^2 \to 0$, but $q_{\mu} \not= 0$. After multiplication
the lhs of (\ref{2.26}) vanishes owing to axial current conservation. In the
r.h.s using the Dirac equations for proton and neutron spinors, we have:

\be
\bar{v}_p (p^{\prime}) \Biggl [2m g_A + q^2 F_2 (q^2) \Biggr ] \gamma_5 v_n
(p),
\label{2.27}
\ee
where $g_A = F_1 (0)$ is the neutron $\beta$-decay coupling constant, $g_A =
1.26$ and $m$ is the nucleon mass (assumed to be equal for proton and
neutron). The only way to avoid the discrepancy with the vanishing lhs of
(\ref{2.26}) is to assume that $F_2(q^2)$ has a pole at $q^2 = 0$:

\be
F_2(q^2)_{q^2 \to 0} = - 2m g_A \frac{1}{q^2}
\label{2.28}
\ee
The pole in $F_2(q^2)$ corresponds to appearance of a massless particle with
pion quantum numbers. The matrix element in (\ref{2.26}) has then the form
(at small $q^2$):

\be
\langle p \mid j^+_{\mu 5} \mid n \rangle = g_A \bar{v}_p (p^{\prime}) \Biggl
(\delta_{\mu \nu} - \frac{q_{\mu} q_{\nu}}{q^2} \Biggr ) \gamma_{\nu}
\gamma_5 v_n (p) ,
\label{2.29}
\ee
where conservation of the axial current is evident. The second term in the
rhs of (\ref{2.29}) can be described by the interaction of the axial current
with the nucleon proceeding through intermediate pion, when the axial
current creates virtual $\pi^+$ and then $\pi^+$ is absorbed by neutron
(Fig.1). The
low energy pion-nucleon interaction can be phenomenologically parametrized
by the Lagrangian

\begin{figure}[tb]
\hspace{58mm}
\epsfig{file=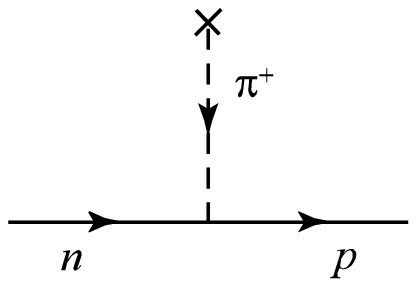}

\vspace{3mm}

{\bf Figure 1.}  The diagram, describing the interaction of nucleon with
axial current through intermediate pion: the solid lines correspond to
nucleon, the dashed line -- to pion, the cross means the interaction with
external axial current.
\end{figure}

\be
L_{\pi NN} = i g_{\pi NN}~ \bar{v}_N \gamma_5 \tau^a v_N \varphi^a
\label{2.30}
\ee
where $\tau^a$ are the isospin Pauli matrices and $g_{\pi NN}$ is the $\pi
NN$ coupling constant, $g^2_{\pi NN}/4 \pi \approx 14$. Using (\ref{2.7})
and (\ref{2.30}) the second term in (\ref{2.26}) can be represented as

\be
- \sqrt{2}~g_{\pi NN} f_{\pi} \bar{v}_p \gamma_5 v_n \frac{q_{\mu}}{q^2}
\label{2.31}
\ee
The comparison with (\ref{2.28}) gives the Goldberger-Treiman relation
\cite{8}

\be
g_{\pi NN} f_{\pi} = \sqrt{2} mg_A
\label{2.32}
\ee
Experimentally, the Goldberger-Treiman relation is satisfied with a 5\%
accuracy, what strongly supports the hypothesis of spontaneous chiral
symmetry violation in QCD. The main modification of (\ref{2.29}) which
arises from the nonvanishing pion mass is the replacement of the pion
propagator: $q^2 \to q^2 - m^2_{\pi}$. Then the contribution of the second
term vanishes at $q_{\mu} \to 0$ and becomes very small in the case of
neutron $\beta$-decay.

Since the only assumption in the consideration above was the conservation of
the axial current, this consideration can be generalized to any other
component of the isospin 1 axial current, if SU(2) flavour symmetry is
supposed, and to any octet axial current in the case of the SU(3) flavor
symmetry. In the last case we come to the conclusion that the octet of
pseudoscalar mesons is massless in the limit of massless $u,d,s$ quarks.

The massless bosons which arise through spontaneous symmetry breaking are
called Goldstone bosons and the theorem which states their appearance is
called Goldstone theorem \cite{9} (see also [11]). The proof of the
Goldstone theorem presented above was based on existence of massive baryons
and on nonvanishing nucleon $\beta$-decay constant $g_A$. Before proceeding
to another proof based on the existence of quark condensate in QCD, let us
formulate some general features of spontaneously broken theories.

Let the Hamiltonian of the theory under consideration be invariant under
some Lie group $G$ , i.e., the group generators $Q_i$ to commute with the
Hamiltonian
\be
 [Q_i, H] = 0, ~~~ i = 1, ...n
\label{2.33}
\ee
The symmetry is spontaneously broken if the ground state is not invariant
under $G$ and a subset of $Q_l,~l \leq m,~1 \leq m \leq n$ exists such that

\be
Q_l \mid 0 \rangle \not= 0
\label{2.34}
\ee
Denote: $\mid B_l \rangle = Q_l \mid 0 \rangle$. As follows from (\ref{2.33})

\be
H \mid B_l \rangle = 0
\label{2.35}
\ee
-- the states $\mid B_l \rangle$ have the same energy as vacuum. These
states may be considered as massless bosons at rest -- Goldstone bosons.
\footnote{The statement that $Q_l$ are generators of a continuous Lie group
is essential -- the theorem is not correct for discrete symmetry generators.}
The generators $Q_j,~ j = m+ 1, ...n$ generate a subgroup $K \subset$G,
since from

\be
Q_j \mid 0 \rangle = 0
\label{2.36}
\ee
it follows
\be
[Q_j, Q_{j^{\prime}}] \mid 0 \rangle = 0~~~~ j, j^{\prime} = m + 1, ... n
\label{2.37}
\ee
In the case of QCD the group $G$ is $SU(3)_L \times SU(3)_R$, which is
spontaneously broken to $SU(3)_V$ -- the group, where generators are the
octet of vector charges. $Q_l$ are the octet of axial charges and $\mid B_l
\rangle$ are the octet of pseudoscalar mesons. (If only $u,d$ quarks are
considered as massless, all said above may be repeated, but relative to
$SU(2)_L \times SU(2)_R$ group).

Strictly speaking, the states $\mid B_l \rangle$ are not well defined, they
have infinite norm. Indeed,

\be
\langle B_l \mid B_l \rangle = \langle 0 \mid Q_l Q_l \mid 0 \rangle = \int~
d^3 x \langle 0 \mid j_l ({\bf x}, t) Q_l (t) \mid 0 \rangle,
\label{2.38}
\ee
where $j_l(x)$ is the charge density operator corresponding to the generator
$Q_l$ .Extracting the ${\bf x}$-dependence of $j_l({\bf x}, t)$ and using
the fact that vacuum and intermediate states in (\ref{2.38}) have zero
momenta, we have

\be
\langle B_l \mid B_l \rangle = \int~ d^3 x \langle 0 \mid j_l (0, t) Q_l(t)
\mid 0 \rangle = V \langle \mid j_l(0,t), ~ Q_l(t) \mid 0 \rangle,
\label{2.39}
\ee
where V is the total volume, $V \to \infty$. Physically, the infinite norm
is well understood, since the massless Goldstone boson with zero momentum is
distributed over the whole space. The prescription
how to treat the problem is evident -- to give a small mass to the boson. In
what follows, when the commutators will be considered, the problem can be
circumvented by performing first the commutation resulting in
$\delta$-functions, and after integration over $d^3 x$.

Let us demonstrate now, how this general theorem works in QCD in a explicit
way. Go back to Eq.(\ref{2.21}), which at $q = 0$ may be rewritten as

\be
\langle 0 \mid [Q^-_5, \bar{u} \gamma_5 d] \mid 0 \rangle =
- \langle 0 \mid \bar{u} u + \bar{d} d \mid 0 \rangle,
\label{2.40}
\ee
where
\be
Q^-_5 = \int~ d^3 x j^-_{0 5} (x)
\label{2.41}
\ee
is the axial charge generator. It is evident from (\ref{2.40}), that $Q^-_5$
does not annihilate vacuum, i.e. it belongs to the set of (\ref{2.34})
generators. It is clear that the same property are inherent to all members
of the octet of axial charges in SU(3) symmetry (or to members of isovector
axial charges in SU(2) symmetry).
Applying the general considerations of Goldstone, Salam and Weinberg
\cite{11} to our case, consider the vacuum commutator

\be
\langle 0\mid [j^- _{\mu 5}(x), \bar{u}(0)\gamma_5 d(0)]\mid 0\rangle
\label{2.42}
\ee
in coordinate space. Eq.(\ref{2.42}) can be written via Lehmann-K\"allen
representation

\be
\langle 0 \mid[j^- _{\mu 5}(x), \bar{u}(0)\gamma_5 d(0)]\mid  0 \rangle =
\frac{\partial}{\partial x_{\mu}} \int
d\kappa^2\Delta(x,\kappa^2)\rho^-(\kappa^2),
\label{2.43}
\ee
where $\Delta(x,\kappa^2)$  is the Pauli-Jordan (causal)  function for a
scalar particle with mass $\kappa$

\be
(\partial^2_{\mu} + \kappa^2) \Delta(x,\kappa^2) = 0
\label{2.44}
\ee
and $\rho(\kappa^2)$ is the spectral function, defined by

\be
(2\pi)^{-3}p_{\mu} \theta(p_0)\rho^-(p^2) = -\sum_n \delta^4(p-p_n)\langle
0\mid j^- _{\mu 5}(0)\mid n \rangle \langle n
\mid \bar{u}(0)\gamma_5 d(0) \mid 0 \rangle
\label{2.45}
\ee
The  axial current conservation and (\ref{2.44}) imply that

\be
\kappa^2\rho^-(\kappa^2) = 0,
\label{2.46}
\ee
hence

\be
\rho^-(\kappa^2) = N\delta(\kappa^2)
\label{2.47}
\ee
The substitution of (\ref{2.47}) into (\ref{2.43}) gives

\be
\langle 0 \mid [j^- _{\mu 5}(x),~\bar{u}(0)\gamma_5 d(0)]\mid  0 \rangle =
\frac{\partial}{\partial x_{\mu}}~ D(x)N,
\label{2.48}
\ee
where $D(x)=\Delta(x,0)$. Put $\mu=0$, $t=0$, integrate (\ref{2.48}) over
$d^3x$  and use the equality $\partial D(x)/\partial
t\mid_{t=0}=-\delta^3(x)$. The comparison of the result with (\ref{2.21})
shows, that $N$ is proportional to quark condensate and nonzero. This means
that the spectrum of physical states contains a massless Goldstone boson
which gives a nonzero contribution to $\rho^-$.
Its quantum numbers
are those of $\pi^+$. It is easy to perform a similar consideration for
other members of the pion multiplet in the case of SU(2) symmetry or for the
pseudoscalar meson octet in the case of SU(3) symmetry.  Obviously, the
proof can be repeated for any other operator  whose commutator with axial
charges has nonvanishing vacuum average.

The presented above two proofs cannot be considered as a rigorous ones, like
a mathematical theorem, where the presence of Goldstone bosons in QCD is
proved starting from QCD  Lagrangian and by use of the first principles of
the theory. Indeed, in the first proof the existence of massive nucleon was
taken as an experimental fact. In the second proof the appearance of
nonvanishing quark condensate in QCD was exploited. The latter was proved
(see eq.'s (\ref{2.20})--(\ref{2.22})) -- basing on Ward identities, which,
as was demonstrated, became selfconsistent only in the case of existence of
massless pion. Therefore, these proofs may be treated as a convincing
physical argumentation, but not a mathematical theorem (cf.[13]).

\section{Nucleon mass and quark condensate}

Let us show now, that the mentioned above two arguments in the favor of
spontaneously broken chiral symmetry in QCD, namely, the existence of
large baryon masses and the appearance of violating chiral symmetry quark
condensate are in fact deeply interconnected. Demonstrate, that baryon
masses arise just due to quark condensate. I will use the QCD sum rule
method invented by Shifman, Vainstein and Zakharov [14], in its applications
to baryons [15]. (For a review and collection of relevant original papers
see [16]). The idea of the method is that at virtualities of order $Q^2 \sim
1$ GeV$^2$ the operator product expansion (OPE) may be used in consideration
of hadronic vacuum correlators. In OPE the nonoperturbative effects reduce to
appearance of vacuum condensates and condensates of the lowest dimension
play the most important role. The perturbative terms are moderate and do not
change the results in essential way, especially in the cases of
chiral symmetry violation, where they can appear as corrections only.

For definiteness consider the proton mass calculation [15,17]. Introduce the
polarization operator

\be
\Pi(p) = i~\int~d^4 x e^{ipx} \langle 0 \vert T { \eta(x), \bar{\eta} (0) }
\vert 0 \rangle \label{(49)}
\ee
where $\eta(x)$ is the quark current with proton quantum  numbers and $p^2$
is chosen to be space-like, $p^2 < 0, ~\vert p^2 \vert \sim 1$ $GeV^2$. The
current $\eta$ is the colourless product of three quark fields, $\eta =
\varepsilon^{abc}~q^a q^b q^c,~ q = u, d$, the form of the current will be
specialized below. The general structure of $\Pi(p)$ is

\be
\Pi(p) = \hat{p} f_1 (p) + f_2(p) \label{(50)}
\ee
The first structure, proportional to $\hat{p}$ is conserving chirality, while
the second is chirality violating.

\begin{figure}[tb]
\hspace{25mm}
\epsfig{file=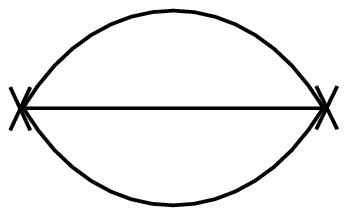}
\hspace{39mm}
\epsfig{file=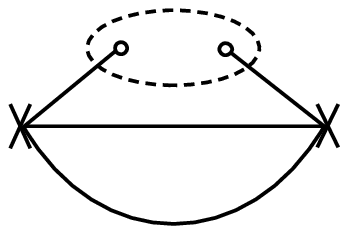}

\vspace{5mm}

\begin{tabular}{p{75mm}p{75mm}}
{\bf Figure 2.}  The bare loop diagram, contributing to chirality
conserving function $f_1(p^2)$: solid lines correspond to quark propagators,
crosses mean the interaction with external currents.  &
{\bf Figure 3.}  The diagram, corresponding to chirality violating dimension
3 operator (quark condensate). The dots, surrounded by circle mean quarks in
the condensate phase. All other notation is the same as on Fig.2.
\end{tabular}
\end{figure}

For each of the functions $f_i(p^2), ~ i = 1,2$ the OPE can be written as:

\be
f_i(p^2) = \sum\limits_{n}~ C^{(i)}_n (p^2) \langle 0 \vert O^{(i)}_n \vert
0 \rangle \label{(51)}
\ee
where $\langle 0 \vert O^{(i)}_n \vert 0 \rangle$ are vacuum expectation
values (v.e.v) of various operators (vacuum condensates), $C^{(i)}_n$ are
coefficient functions calculated in QCD. For the first, conserving chirality
structure function $f_i(p^2)$ OPE starts from dimension zero $(d = 0)$ unit
operator. Its contribution is described by the diagram of Fig.2 and

\be
\hat{p} f_1(p^2) = C_0 \hat{p} p^4 ln [\Lambda^2_u/(-p^2) ] + polynomial,
\label{(52)}
\ee
where $C_0$ is a constant, $\Lambda_u$ is the ultraviolet cutoff. The OPE
for chirality violating structure $f_2(p^2)$ starts from $d = 3$ operator,
and its contribution is represented by the diagram of Fig.3:

\be
f_2(p^2) = C_1 p^2 \langle 0 \vert 0 \bar{q} q \vert 0 \rangle ln
\frac{\Lambda^2_u}{(-p^2)} + polynomial \label{(53)}
\ee
Let us for a moment restrict ourselves to this first order terms of OPE and
neglect higher order terms (as well the perturbative corrections).

On the other hand, the polarizaion operator (49) may be expressed
via the characteristics of physical states using the dispersion relations

\be
f_i(s) = \frac{1}{\pi}~ \int~ \frac{Im f_i (s^{\prime})}{s^{\prime}+s}~
ds^{\prime} + polynomial,~~ s = -p^2  \label{(54)}
\ee

The proton contribution to $Im \Pi(p)$ is equal to

\be
Im \Pi(p) = \pi \langle 0 \vert \eta \vert p \rangle \langle p \vert
\overline{\eta} \vert 0 \rangle \delta(p^2-m^2) = \pi \lambda^2_N (\hat{p}+m)
\delta(p^2-m^2), \label{(55)}
\ee

where

\be
\langle 0 \vert \eta \vert p \rangle = \lambda_N v(p), \label{(56)}
\ee
$\lambda_N$ is a constant, $v(p)$ is the proton spinor and $m$ is the proton
mass. Still restricting ourselves to this rough approximation, we may take
equal the calculated in QCD expression for $\Pi(p)$ (Eq.'s(52),(53)) to its
phenomenological representation Eq.(55). The best way to get rid of
unknown polynomial, is to apply to both sides of the equality the
Borel(Laplace) transformation, defined as

\be
{\cal{B}}_{M^2}f(s) = \lim_{n \to \infty, s \to \infty, s/n=M^2=Const}~
\frac{s^{n+1}}{n!}~ \Biggl( -\frac{d}{ds} \Biggr )^n f(s) = \frac{1}{\pi}~
\int\limits^{\infty}_{0}~ ds Im f(s) e^{-s/M^2} \label{(57)}
\ee
if $f(s)$ is given by dispersion relation (54). Notice, that

\be
{\cal{B}}_{M^2} \frac{1}{s^n} = \frac{1}{(n-1)!(M^2)^{n-1}} \label{(58)}
\ee

Owing to the factor $1/(n-1)!$ in (58) the Borel transformation
suppresses the contributions of high order terms in OPE.

Specify now the quark current $\eta(x)$. It is clear from (55) that
proton contribution will dominate in some region of the Borel parameter $M^2
\sim m^2$ only in the case when both calculated in QCD functions $f_1$ and
$f_2$ are of the same order. This requirement, together with the
requirements of absence of derivatives  and of renormcovariance fixes the
form of current in unique way (for more details see [15,18]):

\be
\eta(x) = \varepsilon^{abc} (u^a C \gamma_{\mu} u^b) \gamma_{\mu} \gamma_5
d^c
\label{59}
\ee
where $C$ is the charge conjugation matrix. With the current $\eta(x)$
(59) the calculations of the diagrams Fig.2 can be easily performed,
the constants $C_0$ and $C_1$ are determined and after Borel transformation
two equations (sum rules) arise (on the phenomenoloical sides of the sum
rules only proton state is accounted)

\be
M^6 = \tilde{\lambda}^2_N e^{-m^2/M^2}
\label{60}
\ee

\be
-2 (2 \pi)^2 \langle 0 \vert \bar{q} q \vert 0 \rangle M^4 = m
\tilde{\lambda}^2_N~e^{-m^2/M^2}
\label{61}
\ee

$$\tilde{\lambda}^2_N = 32\pi^4~\lambda^2_N $$

It can be shown that this rough approximation is valid at $M \approx m$.
Using this value of $M$ and dividing (60) on (61) we get a
simple formula for proton mass [15]:

\be
m = [ - 2 (2 \pi)^2 \langle 0 \vert \hat{q} q \vert 0 \rangle ]^{1/3}
\label{(62)}
\ee
This formula demonstrates the fundamental fact, that the appearance of the
proton mass is caused by spontaneous violation of chiral invariance: the
presence of quark condensate. (Numerically, (62) gives the
experimental value of proton mass with an accuracy better than 10\%).

A more refined treatment of the problem of the proton mass calculation was
performed: high order terms of OPE were accounted, as well as excited states
in the phenomenological sides of the sum rules and the stability of the Borel
mass dependence was checked. In the same way, the hyperons, isobar and some
resonances masses were calculated, all in a good agreement with experiment
[19,20,21]. I will not dwell on these results. The main conclusion is: the
origin of baryon masses is in spontaneous violation of chiral invariance --
the existence of quark condensate in QCD. Therefore, three phenomena: baryon
masses, quark condensate and the appearance of Goldstone bosons are
tightly connected.

\section{Chiral effective theory at low energies}

An effective chiral theory based on QCD and exploiting the existence and
properties of the Goldstone bosons may be formulated. This
theory is an effective low energy theory, what means that the theory is
selfconsistent, but only in terms of expansion in powers of particle momenta
(or in the derivatives of fields in the coordinate space). The Lagrangian is
represented as a series of terms with increasing powers of momenta. The
theory breaks down at sufficiently high momenta, the characteristic
parameters are $\mid {\bf p}_i \mid /M$, where ${\bf p}_i$ are the spatial
momenta of the Goldstone bosons entering the process under consideration and
$M$ is the characteristic scale of strong interaction. (Since ${\bf p}_i$
depend on the reference frame, some care must be taken when choosing the
most suitable frame in each particular case). The physical ground of the
theory is the fact that in the limit of vanishing (or small enough) quark
masses the spectrum of Goldstone bosons is separated by the gap from the
spectrum of other hadrons.  The chiral effective theory working in the domain
$\mid {\bf p}_i \mid /M \ll 1$, is a selfconsistent theory and not a model.
Such theory can be formulated basing on the $SU(2)_L \times SU(2)_R$
symmetry with pions as (quasi) Goldstone bosons. Then one may expect the
accuracy of the theory to be of the order of the one of the isospin theory,
i.e. of a few per cent. Or the theory may be based on the $SU(3)_L \times
SU(3)_R$ symmetry with an octet of pseudoscalar bosons $\pi, K, \eta$ as
(quasi) Goldstone bosons. In this case the accuracy of the theory is of
order of violation of the SU(3) symmetry, i.e., of order $m_s/M \sim
10-20\%$. For definiteness, the main part of this section deals with the
case of $SU(2)_L \times SU(2)_R$.

The heuristic arguments for the formulation of the chiral theory are the
following. In the limit of quark and pion mass going to zero (\ref{2.7})
can be replaced by the field equation

\be
j^i_{\mu 5} = -(f_{\pi}/\sqrt{2}) \partial_{\mu} \varphi^i_{\pi}
\label{2.49}
\ee

\be
j^i_{\mu 5} = \bar{q} \gamma_{\mu} \gamma_5 (\tau^i/2) q, ~ q = u,d
\label{2.50}
\ee
where $\varphi^i_{\pi}$ is the pion field, $\tau^i$ are the Pauli matrices
and $i = 1,2,3$ is isospin index. (Normalization of the current
$j^i_{\mu 5}$ is changed comparing with (\ref{2.7}) in order to have the
standard commutation relations of current algebra). Taking the divergence
from (\ref{2.49}) we have

\be
\partial_{\mu} j^i_{\mu 5} = (f_{\pi}/\sqrt{2})m^2_{\pi} \varphi^i_{\pi}
\label{2.51}
\ee
Eqs.(\ref{2.49}),(\ref{2.51}) are correct near the pion mass shell.

Since the pion state is separated by the gap from the
other massive states in the channel with pion quantum numbers these
equations can be treated as a field equations valid in the low energy region
(usually they are called the equations of partial conservation of axial
current PCAC).

The direct consequence of (\ref{2.51}) is the Adler selfconsistency
condition [22]. Consider the amplitude of the process $A \to B + \pi$, where
A and B are arbitrary hadronic states in the limit of vanishing pion
momentum $p$. The matrix element of this process can be written as

\be
M_i(2 \pi)^4 \delta^4(p_A - p - p_B) = \int~d^4x e^{ipx}
(\partial^2_{\mu} +m^2_{\pi}) \langle
B \mid \varphi^i_{\pi} \mid A \rangle
\label{2.52}
\ee
The substitution of (\ref{2.51}) gives

\be
M_i = \frac{i(p^2-m^2_{\pi})}{(f_{\pi}/ \sqrt{2}) m^2_{\pi}} p_{\mu}
\langle B \mid j^i_{\mu
5}(0) \mid A \rangle
\label{2.53}
\ee
Going to the limit $p_{\mu} \to 0$ we get

\be
M(A \to B\pi)_{p \to 0} \to 0
\label{2.54}
\ee
what is the Adler condition. When deriving (\ref{2.54}) it was implicitly
assumed that the matrix element $\langle B \mid j^i_{\mu 5} \mid A \rangle$
does not contain pole terms, where the axial current interacts with an
external line. Generally, the Adler theorem does not work in such cases.

The chiral theory is based on the following principles:

1. The pion field transforms under some representation of the group $G =
SU(2)_L \times SU(2)_R$.

2. The action is invariant under these transformations.

3. After breaking the transformations reduce to $SU(2)$ -- the
transformations, which are generated by the isovector vector current.

4. In the lowest order the field equations (\ref{2.49}), (\ref{2.51}) are
fulfilled.

The pion field may be represented by the $2 \times 2$ unitary matrix $U(x),
~U^{-1} =  U^+(x)$, depending on $\varphi^i_{\pi}(x)$.  The condition
$det U = 1$ is imposed on $U(x)$. Therefore the number of degrees of
freedom of matrix $U$ is equal to that of three pionic fields
$\varphi^i_{\pi}(x)$. The transformation law under the group $G$
transformations is given by

\be
U^{\prime}(x) = V_L U(x) V^+_R
\label{2.55}
\ee
where $V_L$ and $V_R$ are unitary matrices of $SU(3)_L$ and $SU(3)_R$
transformations. (\ref{2.55}) satisfies the necessary condition, that after
breaking, when $G$ reduces to $SU(2)$ and $V_L = V_R = V$, the
transformation law reduces to

\be
U^{\prime} = V U(x) V^{-1}
\label{2.56}
\ee
-- the transformation, induced by the vector current.

It can be shown that the general form of the lowest order effective
Lagrangian, where only the terms up to $p^2$ are kept and the breaking
arising from the pion mass is neglected, is: [23]-[26]

\be
L_{eff} = k~ Tr (\partial_{\mu} U \cdot \partial_{\mu}U^+),
\label{2.57}
\ee
where $k$ is some constant.

The conserving vector and the axial currents (Noether currents),
corresponding to Lagrangian (\ref{2.57}) can be found by applying to
(\ref{2.57}) the transformations (\ref{2.55}) with

\be
V_L = V_R = 1 + i \vec{\varepsilon} \vec{\tau}/2
\label{2.58}
\ee
in case of vector current and

\be
V_L = V^+_R = 1 + i \vec{\varepsilon} \vec{\tau}/2,
\label{2.59}
\ee
in case of axial current. (Here
$\vec{\varepsilon}$ is an infinitesimal isovector). The results are:

$$
j^i_{\mu} = ik~Tr (\tau_i [\partial_{\mu} U, U^+])
$$

\be
j^i_{\mu 5} = ik~Tr(\tau_i \{ {\partial_{\mu} U, U^+} \})
\label{2.60}
\ee

One may use various realizations of the matrix field $U(x)$ in terms of
pionic fields $\varphi^i_{\pi}(x)$. All of them are equivalent and lead to
the same physical consequences \cite{18, 19}. Mathematically, this is
provided by the statement that one realization differs from the other by a
unitary (nonlinear) transformation (\ref{2.55}). One of the useful
realizations is

\be
U(x) = exp~ (i \alpha \vec{\tau} \vec{\varphi}_{\pi} (x)),
\label{2.61}
\ee
where $\alpha$ is a constant. Substitution of (\ref{2.61}) into (\ref{2.57})
and expansion in power of pionic field up to the 4-th power gives

\be
L_{eff} = 2 k \alpha^2(\partial_{\mu} \vec{\varphi}_{\pi})^2 + \frac{2}{3} k
\alpha^4 \Biggl [(\vec{\varphi}_{\pi} \partial_{\mu} \vec{\varphi}_{\pi})^2
- \vec{\varphi}^2_{\pi} \cdot (\partial_{\mu} \vec{\varphi}_{\pi})^2 \Biggr
] + ...
\label{2.62}
\ee
From the requirement that the first, kinetic
energy, term in (\ref{2.62}) has the standard from, we have

\be
k \alpha^2 = \frac{1}{4}
\label{2.63}
\ee
Substitution of (\ref{2.61}) into (\ref{2.60}) in the first nonvanishing
order in pionic field and account of (\ref{2.63}) results
in:

$$
j^i_{\mu} = \varepsilon_{ikl}~ \varphi^k_{\pi} \frac{\partial
\varphi^l_{\pi}}{\partial x_{\mu}}
$$
\be
j^i_{\mu 5} = - 2 \sqrt{k} \frac{\partial \varphi^i_{\pi}}{\partial x_{\mu}}
\label{2.64}
\ee
The formula for the vector current -- the first in eqs.(\ref{2.64}) is
the standard formula for the pion isovector current.The comparison of
the second equation (\ref{2.64}) with (\ref{2.49}) finally fixes the constant
$k$ and, because of (\ref{2.63}), $\alpha$

\be
k = \frac{1}{8} f^2_{\pi},~ \alpha = \frac{\sqrt{2}}{f_{\pi}}
\label{2.65}
\ee
Therefore, the effective Lagrangian (\ref{2.57}) as well as $U(x)$ are
expressed through one parameter - the pion decay constant $f_{\pi}$, which
plays the role of the coupling constant in the theory. From dimensional
grounds it is then clear that the expansions in powers of the momenta or in
powers of pionic field are in fact the expansions in $p^2/f^2_{\pi}$ and
$\varphi^2/f^2_{\pi}$. Particularly, the expansion of the effective
Lagrangian (\ref{2.62}) takes the form

\be
L_{eff} = \frac{1}{2} (\partial_{\mu} \vec{\varphi}_{\pi})^2 + \frac{1}{3}
\frac{1}{f^2_{\pi}} \Biggl [(\vec{\varphi}_{\pi} \partial_{\mu}\vec
{\varphi}_{\pi})^2 - \vec{\varphi}^2_{\pi}(\partial_{\mu} \vec
{\varphi}_{\tau})^2 \Biggr ] + ...
\label{2.66}
\ee

Turn now to symmetry breaking term in the chiral effective theory
Lagrangian. This term is proportional to the quark mass matrix

\be
{\cal{M}} = \Biggl (
\begin{array}{ll}
m_u & 0\\
0  & m_d
\end{array}
\Biggr )
\label{2.67}
\ee
In the QCD Lagrangian the corresponding term transforms under $SU(2)_L
\times SU(2)_R$ transformations according to representation $\frac{1}{2},
\frac{1}{2}$. This statement may be transferred to chiral theory by the
requirement that in chiral theory the mass matrix (\ref{2.67}) transforms
according to

\be
{\cal{M}^{\prime}} = V_R {\cal{M}} V^+_L
\label{2.68}
\ee
The term in the Lagrangian linear in ${\cal{M}}$ and of the lowest (zero)
order in pion momenta, invariant under $SU(2)_L \times SU(2)_R$
transformation has the form

\be
L^{\prime} = \frac{f^2_{\pi}}{4}\{B Tr ({\cal{M}} U^+) + B^* Tr ({\cal{M}}
U)\}, \label{2.69} \ee where $B$ is a constant and the factor $f^2_{\pi}$ is
introduced for convenience. Impose the requirement of T-invariance to the
Lagrangian (\ref{2.69}). The pion field is odd under $T(\varphi^i_{\pi})=
-\varphi^i_{\pi}$, so $T U = U^+$ and, as a consequence, $B = B^*$ and

\be
L^{\prime} = \frac{f^2_{\pi}}{4} B Tr [{\cal{M}} (U + U^*)]
\label{2.70}
\ee
In the lowest orders of the expansion in pionic fields (\ref{2.70}) reduces
to

\be
L^{\prime} = \frac{1}{2} B(m_u + m_d) \Biggl [f^2_{\pi} -
\vec{\varphi}^2_{\pi} + \frac{1}{6 f^2_{\pi}}(\vec{\varphi}^2_{\pi})^2
\Biggr ] \label{2.71} \ee The first term in the square bracket gives a shift
in vacuum energy resulting from symmetry breaking, the second corresponds to
the pion mass term in the Lagrangian $-(m^2_{\pi}/2) \vec{\varphi}^2_{\pi}$.
With this identification we can determine the constant $B$:

\be
B = \frac{m^2_{\pi}}{m_u + m_d} = -\frac{2}{f^2_{\pi}} \langle 0 \mid
\bar{q} q \mid 0 \rangle,
\label{2.72}
\ee
where the Gell-Mann-Oakes-Renner relation (\ref{2.23}) was used. The
relation (\ref{2.72}) can be also obtained in an another way. We have from
the QCD Lagrangian

\be
\frac{\partial}{\partial m_u} \langle 0 \mid L \mid 0 \rangle = - \langle 0
\mid \bar{u}u \mid 0 \rangle
\label{2.73}
\ee
Differentiating (\ref{2.71}) we get:

\be
\frac{1}{2} B f^2_{\pi} = - \langle 0 \mid \bar{u} u \mid 0 \rangle ,
\label{2.74}
\ee
what coincides with (\ref{2.72}).

As a simplest application of the effective Lagrangians (\ref{2.66}),
(\ref{2.71}), calculate the pion-pion scattering amplitude in the first
order in $1/f^2_{\pi}$. The results are \cite{20}:

\be
T = \delta^{ik} \delta^{lm} A(s,t,u) + \delta^{il} \delta^{km} A(t,s,u) +
\delta^{im} \delta^{kl} A(u,t,s)
\label{2.75}
\ee
where

\be
A(s,t,u) = \frac{2}{f^2_{\pi}}(s - m^2_{\pi})
\label{2.76}
\ee

\be
s = (p_1 + p_2)^2, ~ t = (p_1 - p_3)^2, u = (p_1 - p_4)^2
\label{2.77}
\ee
$p_1, p_2$ - are initial and $p_3, p_4$ are final pion momenta.
The isospin indices $i,k$ refer to initial pions, $l, m$ -- to final ones.
For example, for the $\pi^+ \pi^0 \to \pi^+ \pi^0$ scattering amplitude we
get \cite{20}

\be
T = \frac{2}{f^2_{\pi}} (t - m^2_{\pi}),
\label{2.78}
\ee
where $T$ is related to the c.m. scattering amplitude $f_{c.m.}$ by

\be
f_{c.m.} = \frac{1}{16 \pi} \frac{1}{E} T,
\label{2.79}
\ee
and $E$ is the energy of $\pi^+$ in c.m.system.

The other, instead of (\ref{2.61}), often used realization is

\be
U(x) = \frac{\sqrt{2}}{f_{\pi}} [\sigma(x) + i \vec{\tau}
\vec{\varphi}_{\pi}(x)]
\label{2.80}
\ee
supplemented by the constrain

\be
\sigma^2 + \vec{\varphi}^2 = \frac{1}{2} f^2_{\pi}
\label{2.81}
\ee
It can be shown by direct calculations, that the expressions for effective
Lagrangians up to $\varphi^4$ obtained in this realization coincide with
(\ref{2.66}), (\ref{2.71}) on pion mass shell. In higher orders ($\varphi^6,
\varphi^8$ etc.) the expressions for effective Lagrangians in these two
realizations are different even on mass shell. But, according to general
arguments by Coleman, Wess and Zumino \cite{18}, the physical amplitudes
became to be equal after adding one-particle reducible tree diagrams. Since
the SU(2) group is isomorphic to $O(3)$ the realization (\ref{2.80}) is
equivalent to the one, where the $O(4)$ real four-vector $U_i(x)$, which
satisfies the constrain $U_i U^T_i = 1, i = 1,2,3,4$, is used instead of the
$2\times2$ matrix $U(x)$ \cite{16}.

The chiral effective Lagrangian (\ref{2.57}) is the leading term in the
expansion in pion momenta. The next term of order of $p^4$ consistent with
Lorentz and chiral invariance, parity and $G$-parity symmetry has the general
form \cite{16}

\be
L_{2, eff} = l_1 [Tr (\partial_{\mu} U \partial_{\mu} U^+)]^2 + l_2 Tr
(\partial_{\mu} U \partial_{\nu} U^+) Tr (\partial_{\mu} U \partial_{\nu}
U^+)
\label{2.82}
\ee
where $l_1$ and $l_2$ are constants.
The term of the second order in quark
masses is added to (\ref{2.82}). If spacial momenta of pions in the process
under consideration are close to zero $\mid {\bf p} \mid \ll m_{\pi}$, the
contribution of this term is of the same order as (\ref{2.82}), since $p^2
=m^2_{\pi} \sim (m_u +m_d)$. Its general form is \cite{16}

$$
L^{\prime}_{2, eff} = l_4 Tr~ (\partial_{\mu} U \partial_{\mu} U^+) Tr~
[\chi(U + U^+)] + l_6 \{ Tr~ [\chi(U + U^+)]\}^2
$$
\be
+ l_7\{Tr [i \chi(U - U^+)]\}^2,
\label{2.83}
\ee
where

\be
\chi = 2 B{\cal{M}}
\label{2.84}
\ee

In order to perform the next to leading order calculations in chiral
effective theory it is necessary, besides the (\ref{2.82}), (\ref{2.83})
contribution, to go beyond the tree approximation in the leading order
Lagrangians and to calculate one-loop
contributions arising from (\ref{2.57}), (\ref{2.70}). As can be seen, the
parameter of the expansion is $(1/\pi f_{\pi})^2 \sim (1/500 MeV)^2$ and, as
a rule, small numerical coefficients also arise. Therefore, the n-loops
contribution is suppressed compared by the leading order tree approximation
by the factor $[p^2/(\pi f_{\pi})^2]^n$. Loop integrals are divergent and
require renormalization. Renormalization can be performed in an any scheme
which preserves the symmetry of the theory. These can be dimensional
regularization or a method where finite imaginary parts of the scattering
amplitudes are calculated and the whole amplitudes are reconstructed by
using dispersion relations (an example of such calculation is given
below) or any others. The counter terms arising in loop calculations
(pole contributions at $d \to 4$ in dimensional regularization or
subtraction constants in the dispersion relation approach) are absorbed by
the coupling constants of the next order effective Lagrangian, like $l_1$
and $l_2$ in (\ref{2.82}).Theoretically unknown constants $l_i$ are
determined by comparing with the experimental data.

As a result of loop calculations and of the account of higher order terms in
the effective Lagrangian the coupling constant $f_{\pi}$ entering
(\ref{2.57}), (\ref{2.70}) acquire some contributions and is no more equal
to the physical pion decay constant defined by (\ref{2.70}). For this reason
the coupling constant $f_{\pi}$ in (\ref{2.57}), (\ref{2.70}) should be
considered as a bare  one, $f^0_{\pi}$ which will coincide with the physical
$f_{\pi}$ after accounting of all higher order corrections. A similar
statement refers to the connection between $m^2_{\pi}$ and $m_u + m_d$
(\ref{2.72}). If $B$ is considered as a constant parameter of the theory,
then the relation (\ref{2.72}) is modified by high order terms.
Particularly, in the next to leading order \cite{21}

\be
m^2_{\pi} = \tilde{m}^2_{\pi} \Biggl [1 + c(\mu) \frac{m^2_{\pi}}{f^2_{\pi}}
+ \frac{m^2_{\pi}}{16 \pi^2 f^2_{\pi}}~ ln \frac{m^2_{\pi}}{\mu^2} \Biggr ],
\label{2.85}
\ee
where

\be
\tilde{m}^2_{\pi} = B(m_u + m_d),
\label{2.86}
\ee
$\mu$ is the normalization point and $c(\mu)$ is the $\mu$-depending
renormalized coupling constant expressed through $l_i$. (The total
correction is $\mu$-independent). The appearance of the nonanalytic in
$m^2_{\pi}$ (or $m_q$) term $\sim m^2_{\pi} ln m^2_{\pi}$ -- the so called
"chiral logarithm" -- is a specific feature of the chiral perturbation
theory. The origin of their appearance are infrared singularities of the
corresponding loop integrals.  $f_{\pi}$ also
contains the chiral logarithm \cite{21}:

\be
f_{\pi} = f^0_{\pi} \Biggl [ 1 + c_1(\mu) \frac{m^2_{\pi}}{f^2_{\pi}} -
\frac{m^2_{\pi}}{8 \pi^2 f^2_{\pi}}~ ln \frac{m^2_{\pi}}{\mu^2} \Biggr ]
\label{2.87}
\ee

Let us present two examples of loop calculations.

{\bf 1.} Find the nonanalytical, proportional to $ln m^2_{\pi}$ correction
to pion electric radius [31,32].

 The one-loop contribution to pion formfactor comes from
$\pi \pi$ interaction term in the Lagrangian given by
(\ref{2.66}) and is equal to

\be
i \frac{1}{f^2_{\pi}}~ \int~ \frac{d^4 k_1 d^4 k_2}{(2 \pi)^4} \delta(q + k_1
- k_2)(k_1 + k_2)_{\mu} \frac{1}{k^2_1 - m^2_{\pi}}~ \frac{1}{k^2_2 -
m^2_{\pi}}(p_1 + p_2)(k_1 + k_2)
\label{102}
\ee
Here $p_1$ and $p_2$ are the initial and final pion momenta $q$ is the
momentum transfer, $q^2 < 0, p_1 + q = p_2$. The integral in (\ref{102}) can
be calculated in the following way. Consider the integral

$$
i~\int~ \frac{d^4 k_1 d^4 k_2}{(2 \pi)^4} (k_1 + k_2)_{\mu} (k_1 +
k_2)_{\nu}\frac{1}{k^2_1 - m^2_{\pi}}~ \frac{1}{k^2_2 - m^2_{\pi}}
\delta^4(q + k_1 - k_2) =
$$
\be
= A(q^2) (\delta_{\mu \nu} q^2 - q_{mu} q_{\nu})
\label{103}
\ee
The form of the rhs of (103) follows from gauge invariance. Calculate the
imaginary part of $A(q^2)$ at $q^2 > 0$. We have

$$
Im A(q^2)(\delta_{\mu \nu}q^2 - q_{\mu}q_{\nu}) = -\frac{1}{8 \pi^2}~ \int~
d^4k (2k - q)_{\mu} (2k - q)_{\nu}\delta [(q - k)^2] =
$$
\be
= \frac{1}{48 \pi} (q^2 \delta_{\mu \nu} - q_{\mu} q_{\nu})
\label{104}
\ee
(The pion mass can be neglected in our approximation). $A(q^2)$ is
determined by dispersion relation:

\be
A(q^2) = \frac{1}{\pi} \int \limits^{M^2}_{4m^2_{\pi}} ~ \frac{ds}{s -
q^2}~Im A (q^2) = \frac{1} {48 \pi^2} ~ ln \frac{M^2}{4m^2_{\pi} - q^2}
\label{105}
\ee
(The subtraction term is omitted, $M^2$ is a cutoff). Substitution of
(\ref{103}), (\ref{104}) into (\ref{102}) gives for the correction to the
$\gamma \pi \pi$ vertex

\be
(p_1 + p_2)_{\mu} [F(q^2) - 1] = (p_1 + p_2)_{\mu} \frac{q^2}{48 \pi^2
f^2_{\pi}}~ ln \frac{M^2}{4m^2_{\pi} - q^2},
\label{106}
\ee
where $F(q^2)$ is the pion formfactor. The pion electric radius is defined
by

\be
r^2_{\pi} = 6 \frac{d F(q^2)}{d q^2}
\label{107}
\ee
and its nonanalytical in $m^2_{\pi}$ part is equal to

\be
r^2_{\pi} = - \frac{1}{8 \pi^2 f^2_{\pi}} ln m^2_{\pi}
\label{108}
\ee

{\bf 2.} Quark condensate also become the nonanalytical, proportional
to $m^2_{\pi} ln m^2_{\pi}$ correction [33]. Using
(\ref{2.73}) and (\ref{2.71}) we get

\be
\langle 0\mid \bar{u}u\mid 0 \rangle = -\frac{1}{2}f^2 {\pi}B\langle 0 \mid
1 - \frac{\varphi^2_i}{f^2_{\pi}}\mid 0 \rangle
\label{109}
\ee
The mean vacuum value of $\varphi^2_i$ is given by

$$
lim_{x \to 0} \langle 0 \mid T {\varphi_i(x), \varphi_i(0) } \mid 0 \rangle =
\frac{3 i}{(2 \pi)^4}~ lim_{x \to 0} \int~ d^4 k \frac{e^{ikx}}{k^2 -
m^2_{\pi}}=
$$
\be
= A m^2_{\pi} + C m^2_{\pi}ln m^2_{\pi} + ...
\label{110}
\ee

In order to find $C$ differentiate (\ref{110}) over $m^2_{\pi}$ . We have

\be
\frac{3 i}{(2 \pi)^4}~ \int~ d^4 k \frac{1}{(k^2 - m^2_{\pi})^2} = - \frac{3
\pi^2}{(2 \pi)^4} ln~ \frac{M^2}{m^2_{\pi}}
\label{111}
\ee
Substitution of (\ref{111}) into (\ref{109}) with the account of (\ref{2.72})
gives

\be
 \langle 0 \mid \bar{u}u \mid 0 \rangle = \langle 0 \mid \bar{u}u \mid 0
 \rangle_0 \Biggl (1 + \frac{3 m^2_{\pi}}{16 \pi^2 f^2_{\pi}} ~ ln
 \frac{M^2}{m^2_{\pi}} + A m^2_{\pi} \Biggr )
 \label{112}
 \ee

Generalization for three massless quark case, when $s$-quark is also
considered as massless and the symmetry of the Lagrangian is $SU(3)_L \times
SU(3)_R$ is straightforward. The matrix $U(x)$ is $3 \times 3$ unitary
matrix, the leading order Lagrangian has the same forms (\ref{2.57}),
(\ref{2.70}) with an evident difference that the quark mass matrix
${\cal{M}}$ is now $3 \times 3$ matrix. In the formulae for axial and vector
currents (\ref{2.50}), (\ref{2.60}) $\tau_i$ should be substituted by the
Gell-Mann matrices $\lambda_n, n = 1, ...8$ and the same substitution must
be done in the exponential realization of $U(x)$:

\be
U(x) = exp \Biggl (i \frac{\sqrt{2}}{f_{\pi}} \sum \limits_{n}
\lambda_n \varphi_n(x) \Biggr ) \label{2.88}
\ee
where $\varphi_n(x)$ is the
octet of pseudoscalar mesonic fields. Because the algebra of $\lambda_n$
matrices differs from that of $\tau_i$ and, particularly, the anticommutator
${\lambda_n, \lambda_m}$ does not reduce to $\delta_{nm}$, the linear
realization as simple as (\ref{2.80}) is impossible in this case.

The symmetry breaking Lagrangian (\ref{2.70}) in the order of $\varphi^2_n$
-- the mass term in the pseudoscalar meson  Lagrangian - is nondiagonal in
mesonic fields: the effective Lagrangian contains the term proportional to
$(m_u - m_d) A \varphi_3 \varphi_8$. The presence of this term means that the
eigenstates of the Hamiltonian, $\pi^0$ and $\eta$ mesons, are not
eigenstates of $Q^3$ and $Q^8$ generators of $SU(3)_V$: in $\eta$ there is
an admixture of the isospin 1 state (the pion) and vice versa
[34,35].

In general we can write:

\be
H = \frac{1}{2}\tilde{m}^2_{\pi} \varphi^2_3  + \frac{1}{3}
\tilde{m}^2_{\eta} \varphi^2_8 + A(m_u-m_d) \varphi_3 \varphi_8 +
\mbox{kinetic} ~~\mbox{terms},
\label{114}
\ee
The physical $\pi$ and $\eta$  states arises after orthogonalization of the
Hamiltonian (\ref{114})

$$\mid \pi \rangle = cos ~\theta \mid \varphi_3 \rangle - sin~\theta \mid
\varphi_8 \rangle$$

\be
\mid \eta \rangle = sin ~\theta \mid \varphi_3 \rangle + cos~\theta \mid
\varphi_8 \rangle
\label{115}
\ee
It can be shown [33,34,26], that the constant $A$ in (\ref{114}) is equal

\be
A = \frac{1}{\sqrt{3}}~\frac{m^2_{\pi}}{m_u+m_d}
\label{116}
\ee
and the mixing angle is given by (at small $\theta$)

\be
\theta  =
\frac{1}{\sqrt{3}}~\frac{m^2_{\pi}}{m^2_{\eta}-m^2_{\pi}}~\frac{m_u-m_d}{m_u+m_d}
\label{117}
\ee
This result is used in consideration of many problems, where isospin is
violated, e.q. the decay rate $\psi^{\prime} \to J/\psi \pi^0$ [36], the
amplitude of $\eta \to \pi^+ \pi^- \pi^0$  decay. The violating isospin
amplitude $\eta \to \pi^+\pi^-\pi^0$ is found to be [37,38]  (in its
derivation (Eq.(117)) was exploited):

\be
T_{\eta \to \pi^+ \pi^-\pi^0} = \frac{\sqrt{3}}{2 f^2_{\pi}}~ \frac{m_u -
m_d}{m_s - (m_u + m_d)/2} \Biggl (s - \frac{4}{3} m^2_{\pi}  \Biggr )
\label{118}
\ee
where $s = (p_{\eta} - p_{\pi^0})^2$.

In the three flavor case the next to leading
Lagrangian contains few additional terms in comparison with (\ref{2.82}),
(\ref{2.83}) \cite{13,17}

$$
L^{\prime}_{2 eff} = l_3~ Tr~ (\partial_{\mu}U \partial_{\mu} U^+
\partial_{\nu} U \partial_{\nu} U^+) + l_5 ~ Tr~ [\partial_{\mu} U
\partial_{\mu} U^+ \chi (U + U^+)]+
$$
\be
+ l_8~ Tr~ (\chi U \chi U^+ + U \chi^+ U \chi)
\label{2.89}
\ee

In the case of three flavours in the order of $p^4$,
  the term of different origin proportional to the totally antisymmetric
tensor $\varepsilon_{\mu \nu \lambda \sigma}$ arises. As was pointed out by
Wess and Zumino, \cite{22} its occurrence is due to anomalous Ward identities
for vector and axial nonsinglet currents. Witten \cite{23} had presented the
following heuristic argument in the favor of this term. The leading and next
to leading Lagrangians (\ref{2.57}), (\ref{2.70}), (\ref{2.82}),
(\ref{2.83}), (\ref{2.89}) are invariant under discrete symmetries $U(x)
\to U^+(x), U({\bf x}, t) \to U(-{\bf x}, t)$. According to (\ref{2.61})
this is equivalent to $\varphi_i(x) \to - \varphi_i(x)$. In the case of
pions this operation coincides with $G$- parity, but for the octet of
pseudoscalar mesons this is not the case. Particularly, such symmetry
forbids the process $K^+K^- \to \pi^+ \pi^- \pi^0$ and $\eta \pi^0 = \pi^+
\pi^- \pi^0$, which are allowed in QCD. In QCD the symmetry under the sign
change of pseudoscalar meson fields is valid only if supplemented by space
reflection, i.e.  $\varphi_i(-{\bf x}, t) \to -\varphi_i({\bf x}, t)$.
Therefore, one may add to chiral lagrangian a term, which is invariant under
the latter operation, but violates separately ${\bf x} \to -{\bf x}$ and
$\varphi_i(x) \to - \varphi_i(x)$. Evidently, such term is proportional to
$\varepsilon_{\mu \nu \lambda \sigma}$. The general form of the term added
to the equation of motion is unique:

\be
\frac{1}{8}f^2_{\pi} (-\partial^2_{\mu}U^+ +U^+ \partial^2_{\mu}U\cdot
U^+)+\lambda \varepsilon_{\mu\nu\lambda \sigma} \{U^+\partial_{\mu}U\cdot
U^+\partial_{\nu}U\cdot U^+\partial_{\lambda}U\cdot
U^+\partial_{\sigma}U\}=0,
\label{2.90}
\ee
where $\lambda$ is a constant. (Other nonleading terms are omitted).
Eq.(\ref{2.90}) is noninvariant under $U^+ \to U$ and ${\bf x} \to -{\bf x}$
separately, but conserves parity. However, (\ref{2.90})  cannot derived from
local Lagrangian in four dimensional space-time, because the trace of the
second term in the lhs of (\ref{2.90}) vanishes. Witten \cite{23} had shown
that the Lagrangian can be represented formally as an integral over some
five-dimensional manifold, where Lagrangian density is local. The integral
over this manifold reduces to its boundary, which is precisely 4-dimensional
space-time. In the first nonvanishing order in mesonic fields the
contribution to the Lagrangian (the so called Wess-Zumino term \cite{22}) is
equal to: [39-41]

 \be
 \Lambda_{WZ}(U) = n\frac{1}{15\pi^2f^2_{\pi}}\int
d^4x~\varepsilon_{\mu\nu\lambda \sigma}
Tr (\Phi \partial_{\mu} \Phi \partial_{\nu} \Phi \partial_{\lambda}
\Phi \partial_{\sigma}\Phi),
\label{2.91}
\ee
where $\Phi = \sum~ \lambda_m \varphi_m$. The coefficient $n$ in
(\ref{2.91}) is an integer number \cite{23}. This statement follows from the
properties of mapping of 4-dimensional space-time into $SU(3)$ manifold
produced by the field $U$. It is clear from (\ref{2.91}) that $L_{WZ} = 0$
in the case of two flavors: the only antisymmetrical tensor in flavor
indices is $\varepsilon^{ikl}$ and it is impossible to construct
antisymmetrical in coordinates expression from the derivatives of pionic
fields.

In order to find the value of $n$ it is instructive to consider the
interaction with electromagnetic field. In this case the Wess-Zumino
Lagrangian is supplemented by terms which form together with (\ref{2.91}) a
gauge invariant Lagrangian \cite{23}

$$L_{WZ}(U,A_{\mu})=L_{WZ}(U) - en \int d^4x
A_{\mu}J_{\mu}+\frac{ie^2n}{24\pi^2}\int d^4
x\varepsilon_{\mu\nu\lambda\sigma}(\partial_{\mu}A_{\nu})A_{\lambda}\times$$

\be
\times Tr[e^2_q(\partial_{\sigma}U)U_+ e^2_qU^+(\partial_{\sigma}U) +
e_qUe_qU^+(\partial_{\sigma} U)U^+],
\label{2.92}
\ee
where
$$J_{\mu}=\frac{1}{48\pi^2}\varepsilon_{\mu\nu\lambda\sigma}Tr[e_q(
\partial_{\nu}U\cdot U^+)(\partial_{\lambda}U\cdot
U^+)(\partial_{\sigma}U\cdot U^+) +$$

\be
+e_q(U^+\partial_{\nu}U)(U^+\partial_{\lambda} U)(U^+\partial_{\sigma}U)],
\label{2.93}
\ee
$e_q$ is the matrix of quark charges, $e_q = diag(2/3, -1/3, -1/3)$ and $e$
is the proton charge. The amplitude of $\pi^0 \to \gamma \gamma$ decay can
be found from the last term in (\ref{2.92}). It is given by

\be
 T(\pi^0\to \gamma\gamma) = \frac{ne^2}{48\sqrt{2}\pi^2
f_{\pi}} \varepsilon_{\mu\nu\lambda\sigma} F_{\mu\nu}F_{\lambda \sigma}
\label{2.94}
\ee
On the other side, the same amplitude is determined in QCD by anomaly. Use
the anomaly condition \cite{25}-\cite{27}

\be
\partial_{\mu}j^3_{\mu 5}=\frac{\alpha}{2\pi}N_c
(e^2_u-e^2_d)F_{\mu\nu}\tilde{F}_{\mu\nu} = \frac{\alpha}{12\pi}N_c
\varepsilon_{\mu\nu\lambda\sigma} F_{\mu\nu}F_{\lambda \sigma},
\label{2.95}
\ee
where $N_c$ is the number of colors and $e_u, e_d$ are $u$ and $d$ quark
charges. For the amplitude $T(\pi^0 \to \gamma \gamma)$ we have, exploiting
the PCAC condition (\ref{2.51}):

\be
T(\pi^0 \to \gamma\gamma) = \frac{e^2}{48\sqrt{2}\pi^2 f_{\pi}} N_c
\varepsilon_{\mu\nu\lambda\sigma} F_{\mu\nu}F_{\lambda \sigma}
\label{2.96}
\ee
(\ref{2.94}) coincides with (\ref{2.95}), if $n = N_c$ \cite{23}. The other
physically interesting object, the $\gamma \pi^+ \pi^- \pi^0$ vertex is
determined by the second term in the rhs of (\ref{2.92}) and is equal to

\be
\Gamma(\gamma \pi^+\pi^-\pi^0) = -\frac{1}{3}ie
\frac{n}{\pi^2\sqrt{2}f^3_{\pi}}
\varepsilon_{\mu\nu\lambda\sigma} A_{\mu}\partial_{\nu}\pi^+
\partial_{\lambda}\pi^-\partial_{\sigma}\pi^0
\label{2.97}
\ee
Again, if $n = N_c$, this result agrees with QCD calculations based on VAAA
anomaly or with the phenomenological approach, where the anomaly was taken
as granted \cite{28}-\cite{30}.

The chiral effective theory is valid also for the pion-baryon low energy
interaction, where a lot of results was obtained. We restrict ourselves here
to presenting of effective pion-nucleon interaction Lagrangian in the
leading order (see e.g.[1], a good review, where high order terms are
considered is in [48]):

\be
L_{\pi N} = -\frac{g_A}{f_{\pi}\sqrt{2}}
\bar{\psi}_N\gamma_{\mu}\gamma_5 \vec{\tau}
\partial_{\mu}\vec{\varphi}\psi_N - \frac{1}{2f^2_{\pi}} \bar{\psi}_N
\gamma_{\mu}
\vec{\tau}[\vec{\varphi}\partial_{\mu}\vec{\varphi}]
\psi_N,
\label{2.98}
\ee
where $\psi_N$ are nucleon spinors and $g_A$ is the axial neutron
$\beta$-decay constant, $g_A = 1.26$. The first term in (\ref{2.98}) is a
standard pion-nucleon interaction with pseudovector coupling, the second one
represents the contact $\pi \pi N \bar{N}$ interaction.

\section{Low energy sum rules in CET}

Using CET technique important low energy sum rules can be derived, which of
course, are valid also in QCD. The most interesting, which are tested by
experiment, refer to the difference of the polarization operators of vector
and axial currents. Let us define

$$\Pi^U_{\mu\nu} (q) = i\int d^4 x e^{iqx} \langle 0 \mid T\{
U_{\mu}(x),~U_{\nu}(0)^+\} 0 \rangle  =$$
\be
=(q_{\mu}q_{\nu} - q^2
\delta_{\mu\nu})\Pi^{(1)}_U(q^2)
+q_{\mu}q_{\nu} \Pi^{(0)}_U (q^2)
\ee
where

\be
U=V,A ~~~~~V_{\mu} = \overline{u}\gamma_{\mu}d,~~~~~A_{\mu} =
\overline{u}\gamma_{\mu} \gamma_5 d,
\ee
$V_{\mu}4$ and $A_{\mu}$ are vector and axial quark currents. The imaginary
parts of the correlators are the so-called spectral functions $(s = q^2)$:

\be
v_1(s)/a_i(s) = 2 \pi~ Im~ \Pi^{(1)}_{V/A} (s), ~~ a_0(s) = 2m Im
\Pi^{(0)}_A(s),
\label{131}
\ee
which are measured in $\tau$-decay. (Isotopically related to $v_1$ spectral
function is measured in $e^+e^-$- annihilation). The spin $0$ axial spectral
function $a_0(s)$ which is mainly saturated by one pion state will not be
interesting for us now.

$\Pi^{(1)}_V(s)$ and $\Pi^{(1)}_A(s)$ are analytical functions of $s$ in the
complex $s$-plane with a cut along the right semiaxes, starting from the
threshold of the lowest hadronic state: $4 m^2_{\pi}$ for $\Pi^{(1)}_V$ and
$9 m^2_{\pi}$ for $\Pi^{(1)}_A$.  Besides the cut, $\Pi^{(1)}_a(q^2)$ has a
kinematical pole at $q^2 = 0$. This is a specific feature of QCD and CET,
which follows from the chiral symmetry in the limit of massless $u,
d$-quarks and its spontaneous violation. In this limit axial current is
conserved and a massless pion exists. Its contribution to the axial
polarization operator is given by

\be
\Pi^A_{\mu \nu} (q)_{\pi} = f^2_{\pi} \Biggl (\delta_{\mu \nu} -
\frac{q_{\mu} q_{\nu}}{q^2} \Biggr )
\ee
When the quark masses are taken into account, then in the first order of
quark masses, or, what it is equivalent, in $m^2_{\pi}$ Eq.132 is modified
to:

\be
\Pi^A_{\mu \nu}(q)_{\pi} = f^2_{\pi} \Biggl (\delta_{\mu \nu} -
\frac{q_{\mu} q_{\nu}}{q^2 - m^2_{\pi}} \Biggr )
\ee
Decompose (133) in the tensor structures of (129)

\be
\Pi^A_{\mu \nu} (q)_{\pi} = - \frac{f^2_{\pi}}{q^2} (q_{\mu} q_{\nu} -
\delta_{\mu \nu} q^2) - \frac{m^2_{\pi}}{q^2}~q_{\mu} q_{\nu} ~
\frac{f^2_{\pi}}{q^2 - m^2_{\pi}}
\ee
The pole in $\Pi^A_1(q^2)$ at $q^2 = 0$ is evident.

Let us write dispersion relation for $\Pi^V_1(s) - \Pi^A_1(s)$. This may be
nonsubtracted dispersion relation, since perturbative terms (besides the
small contribution from $u, d$ quarks mass square) cancels in the
difference, and the OPE terms decrease with $q^2 = s$ at least as $s^{-2}$
(the term $\sim m_q \langle0 \vert \bar{q} q \vert 0 \rangle$ in OPE). We
have

\begin{figure}[p]
\hspace{33mm}
\epsfig{file=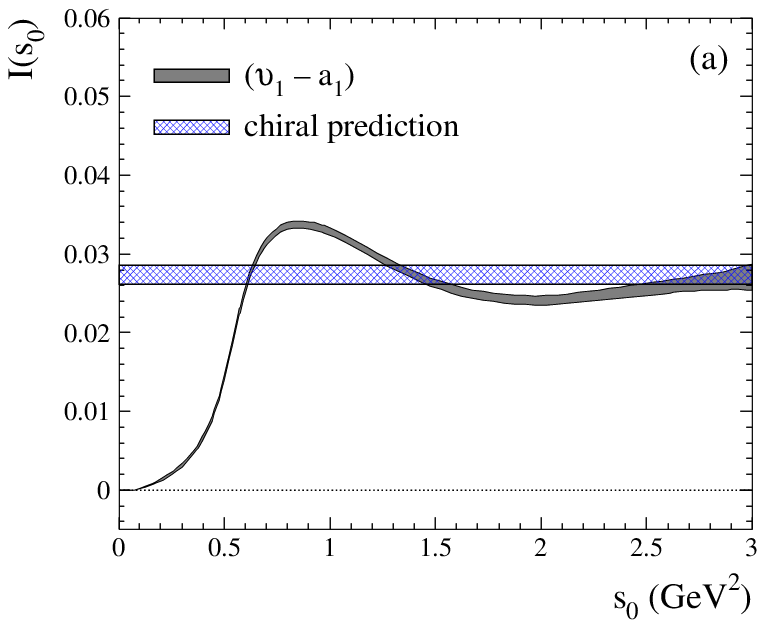}

\hspace{33mm}
\epsfig{file=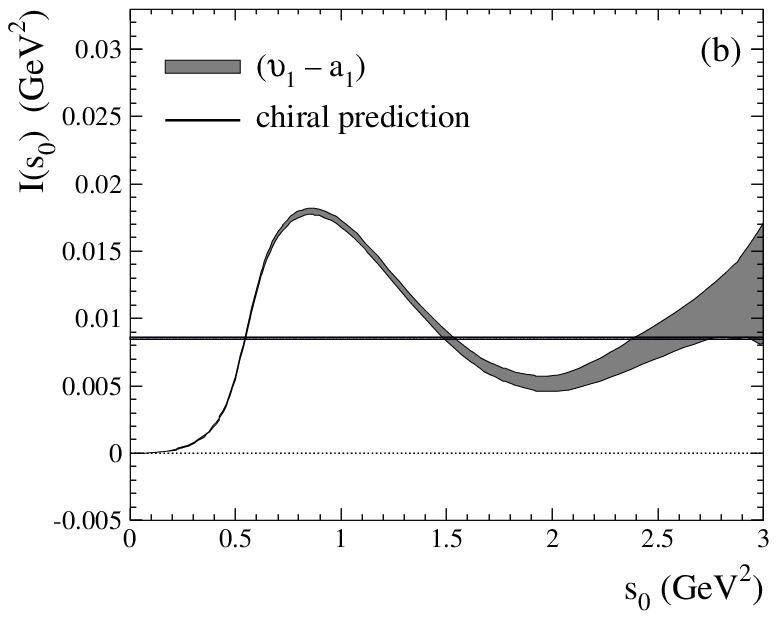}

\hspace{33mm}
\epsfig{file=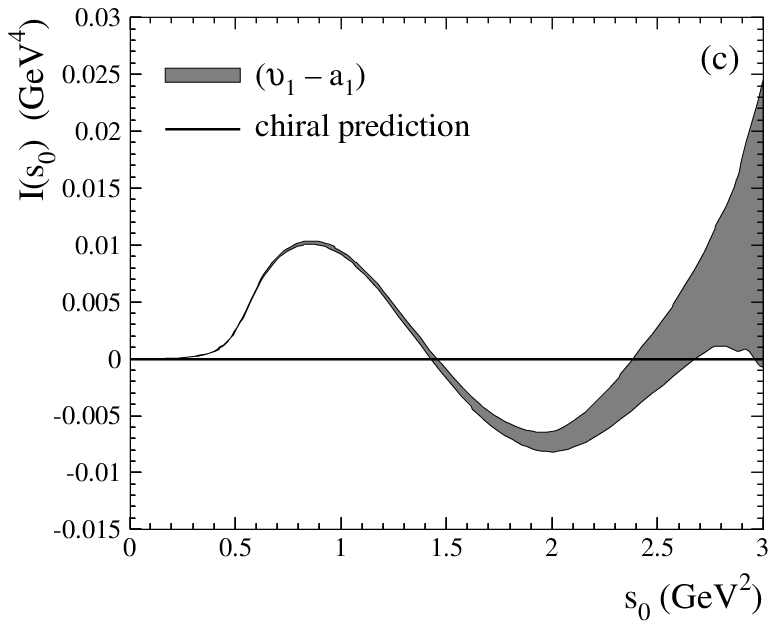}

{\bf Figure 4a,b,c.} The sum rules (136), (137), (138) correspondingly as
functions of upper limits of integration $s_0$. (Data of ALEPH [52])
\end{figure}

\be
\Pi^V_1(s) - \Pi^A_1(s) = \frac{1}{2 \pi^2}~ \int\limits^{\infty}_{0}~
ds^{\prime}~ \frac{v_1(s^{\prime}) - a_1(s^{\prime})}{s^{\prime}-s} +
\frac{f^2_{\pi}}{s}
\ee

The last term in the rhs of (135) represents the kinematical pole
contribution. Let us go to $s \to \infty$ in (135). Since $\Pi^V_1(s) -
\Pi^A_1(s) \to s^{-2}$ in this limit we get the sum rule (the first Weinberg
sum rule [49]):

\be
\frac{1}{2 \pi^2}~ \int\limits^{\infty}_{0} ~ ds [v_1(s) - a_1(s)] =
 f^2_{\pi}
\ee
The accuracy of this sum rule is of order of chiral symmetry violation in
QCD, or next order terms in CET, i.e. $\sim m^2_{\pi}/M^2$ (e.g. a
subtraction term).

If the term $\sim m_q \langle 0 \vert \bar{q} q \vert 0 \rangle~~ \sim
f^2_{\pi} m^2_{\pi}$ in OPE may be neglected, then, performing in (135) the
expansion up to $1/s^2$ we get the second Weinberg sum rule:

\be
\int\limits^{\infty}_{0}~ sds [v_1(s) - a_1(s)] = O (m^2_{\pi})
\ee
(For other derivations of these sum rules -- see [50]).

I present here one more sum rule derived in CET (in its earlier version --
PCAC):

Das-Mathur-Okubo sum rule [51]:

\be
\frac{1}{4 \pi^2}~ \int\limits^{\infty}_{0}~ds \frac{1}{s} [v_1(s) - a_1(s)]
= \frac{1}{6}~f^2_{\pi} \langle r^2_{\pi}\rangle  - F_A,
\ee
where $\langle r^2_{\pi} \rangle$ is the mean pion electromagnetic radius
and $F_A$ is the pion axial vector formfactor in the decay $\pi^- \to m^-
\nu_{\mu} \gamma$ (in fact, $F_A$ is a constant with high accuracy).

The comparison of the sum rules (135, 136, 137) with the results of the
recent measurements of $v_1(s) - a_1(s)$ in $\tau$-decay by ALEPH
collaboration [52] are presented in Fig.4 versus the upper limit of
integration.

\section{QCD and CET at finite temperature}

CET is a useful tool for study QCD at finite temperature. It is a common
believe, that with temperature increase any hadronic system undergoes a
phase transition with restoration of chiral symmetry and liberation of
colors -- deconfinement (for reviews see [53-55]. These two phenomena can
proceed in a single phase transition or may be separated. The estimation of
the critical temperature(s) $T_c$ were found from lattice calculations,
from studies of suitable correlation functions, in the framework of models
and from the study of temperature dependence of condensate in the framework
of CET. All of this indicates that $T_c \approx 150-250 MeV$.

I present here the simple argument [56], based on the consideration on any
hadronic correlator $P(x)$ at large space-like distances ${\bf x}$. One may
expect, that

\be
P(x) \sim e^{-\mu(T) \vert {\bf x} \vert}
\ee
where $\mu(T)$ is temperature depending screening parameter. Eq.139 is
valid if: 1) $\mu \vert {\bf x} \vert \gg 1;~ 2) \vert {\bf x} \vert \la
(\alpha_s(T) T)^{-1}$, because at such $\vert {\bf x} \vert$ the infrared
divergence arises in the theory [57]. At high temperature $\mu(T)$ is given
by Matsubara frequency

\be
\mu = 2 \pi T ~ \mbox{for bosons (two quarks)}, ~~~ \mu = 3 \pi T \mbox{for
baryons (three quarks)}
\ee

At low $T~~ \mu(T)$ is equal by the mass of the corresponding hadron (except
for pion, where the conditions 1 and 2 cannot be satisfied simultaneously).
Taking, as examples, $\rho$ and $a_1$-mesons, we find that the matching of
two regimes occurs at 150-200 MeV.

Quark condensate may be considered as an
order parameter in QCD. Its vanishing at some critical temperature $T = T_c$
would indicate on the phase transition -- the restoration of chiral symmetry
at $T = T_c$. Taking this in mind, calculate
$T^2$
correction to $\langle 0 \mid \bar{u}u \mid 0 \rangle = \langle 0 \mid
\bar{d}d \mid 0 \rangle$ quark condensate in the limit of massless $u, d$
quarks [58,59].

The mean value of any operator $O$  at finite temperature is
given by

\be
\langle O \rangle_T = \sum \limits_{n} \langle n \mid O \frac{1}{e^{H/T} \pm
1} \mid n \rangle \rho_n,
\ee
where $\pm$ signs refer to Fermi and Bose systems, $\rho$ is the density of
the state $\mid n \rangle$. At low $T$ and massless $u, d$ quarks the main
contribution comes from states of massless pions. Contributions of all other
particles are exponentially suppressed by factors $e^{-m/T}$ where $m$ is
the particle mass. (Summation over $n$ should be performed over Hilbert
space of physical particles, since  at small $T$ the system is in
confinement phase and the problem is characterized by large distances). In
the order of $T^2$ it is enough to account in (140) only one pion state.
This gives

\be
\Delta_T \langle \bar{u}u \rangle= 3~ \int~ \frac{d^3 p}{(2 \pi)^3 \cdot 2E}
\langle \pi^+ \mid \bar{u}u \mid \pi^+ \rangle \frac{1}{e^{E/T} - 1}
\ee
where $\Delta_T$ means the temperature correction and factor $3$ comes from 3
pion states -- $\pi^+, \pi^-, \pi^0$. It is clear that the one-pion phase
space factor results in required power $T^2$, two-pion states give $T^4$
etc. From QCD Lagrangian we have

\be
\langle \pi^+ \mid \bar{u}u \mid \pi^+ \rangle = -\frac{\partial}{\partial
m_u} \langle \pi^+ \mid L \mid \pi^+ \rangle
\ee
Substitution of the chiral effective Lagrangian (\ref{2.71}) into (143)
instead of the QCD Lagrangian leads to

\be
\langle \pi^+ \mid \bar{u}u \mid \pi^+ \rangle = \frac{1}{2} B \langle \pi^+
\mid 2 \varphi^+ \varphi \mid \pi^+ \rangle = B = -\frac{2}{f^2_{\pi}}
\langle 0 \mid \bar{u}u \mid 0 \rangle
\ee
Therefore,

\be
\Delta_T \langle \bar{u}u \rangle = -\frac{6}{f^2_{\pi}} \langle 0
\mid \bar{u}u \mid 0 \rangle ~ \int~ \frac{d^3 p}{(2 \pi)^3 \cdot 2E}~
\frac{1}{e^{E/T - 1}}=
- \frac{T^2}{4 f^2_{\pi}} \langle 0 \mid \bar{u}u \mid 0 \rangle
\ee
Quark condensate decreases with increasing of temperature. If such linear
with $T^2$ behaviour would continue up to $T = 2f_{\pi} = \simeq 250 MeV$,
quark condensate would vanish at this temperature and chiral symmetry would
be restored. In fact, the calculation of higher order terms in $T^2$ (up to
$T^6$) gives [60,61]

\be
\langle \bar{q}q \rangle_T = \langle 0 \mid \bar{q}q \mid 0 \rangle \Biggl [
1 - \frac{N^2_f - 1}{N_f}~ \frac{T^2}{6 f^2_{\pi}} - \frac{N^2_f - 1}{2
N^2_f} \Biggl (\frac{T^2}{6 f^2_{\pi}} \Biggr )^2 - N_f (N^2_f - 1) \Biggl (
\frac{T^2}{6 f^2_{\pi}} \Biggr )^3 \cdot ln \frac{M}{T} \Biggr ],
\ee
where $N_f$ is the number of flavors ($N_f = 2$  for $u,d$ massless quarks)
 and $M$ is a cutoff. All three terms in the expansion have the same sign
 what indicates lowering phase transition temperature, up to $T_c \sim 150
 MeV$.
Quark condensate temperature dependence at low $T$  is shown on Fig.5.

\begin{figure}[tb]
\hspace{33mm}
\epsfig{file=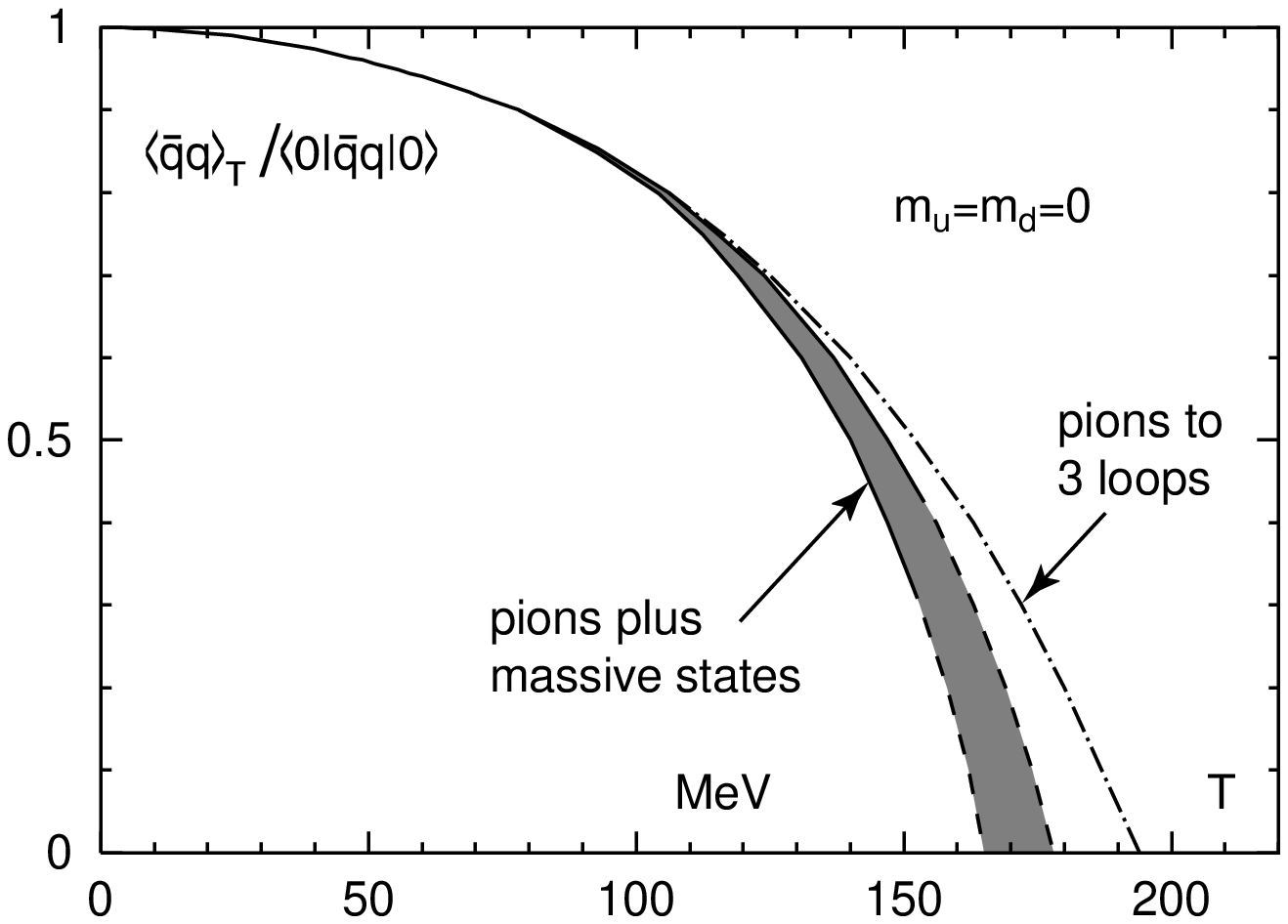, width=90mm}

{\bf Figure 5.}
Temperature dependence of quark condensate up to 3 loops
(at $m_u=m_d=0$)  -- Eq.146 -- dot--dashed line. The shaded area is the same
with model account of massive states (from Ref.61).
\end{figure}

For gluonic condensate the situation is more subtle. The operator $G_{\mu
\nu} G_{\mu \nu}$ is proportional to the trace of the energy-momentum tensor
$\theta_{\mu \nu}$ and the latter is generator of conform
transformation. However, the massless non-interacting pion gas is
conformally invariant. (Pions are non-interacting at low $T$ because of the
Adler theorem). For this reason the low temperature expansion for gluonic
condensate starts from $\sim T^8$ term [60].

Consider finally $T^2$ corrections to the correlators of vector and axial
currents in the limit of massless quarks [62]. At finite $T$ the correlators
are defined as ($q^2 = -Q^2 < 0$).

\be
\Pi^V_{\mu \nu}(q, T) = i~\int~ d^4 x e^{iqx}~\sum\limits_{n}~ \langle n
\vert T {V^a_{\mu}(x),  V^a_{\nu}(0)} exp [(\Omega - H)/T] \vert n \rangle
\ee
where
 \be
 V = V, A~~~ V^a_{\mu} = \bar{q} \gamma_{\mu} \frac{\tau^a}{2}~q, ~~
 A^a_{\mu} = \bar{q}\gamma_{\mu} \gamma_5 \frac{\tau_a}{2} q
 \ee
 and $e^{-\Omega/T} = \sum\limits_{n}~ \langle n \vert e^{-H/T} \vert n
 \rangle$. To evaluate $\Pi^V_{\mu \nu}(q, T)$ at low temperature, $T^2 \ll
 Q^2$ in the sum over $\vert n \rangle$ in (147) only the vacuum and pion
 states must be taken into
account. The matrix elements

\be
\langle \pi \mid T \{ U^a_{\mu} (x),~U^a_{\nu} (0)\} \mid \pi \rangle
\ee
are easily evaluated applying reduction formulas to pions and using Eq.(65).
Equal time commutators, which arise, are calculated by current algebra
relations (or can be derived from Eq.78). The integration over pionic phase
space (relativistic Bose gas)  can be done with the help of the formula (for
massless pions):

\be
\int \frac{d^3 k}{(2\pi)^3}~\frac{1}{2k}~\frac{1}{e^{k/T}-1} = \frac{1}{24}
T^2
\ee
The result is:

$$\Pi^V_{\mu\nu} (q,T) = (1-\varepsilon)\Pi^V_{\mu\nu} (q,0) + \varepsilon
\Pi^A_{\mu\nu} (q,0)$$

\be
\Pi^A_{\mu\nu} (q,T) = (1-\varepsilon)\Pi^A_{\mu\nu} (q,0) + \varepsilon
\Pi^V_{\mu\nu} (q,0)
\ee
where $\varepsilon = T^2/3f^2_{\pi}$. If $\Pi^{V/A}_{\mu\nu}(q,0)$  are
represented through dispersion relations by contributions of physical states
in $V$ and $A$  channels, say $\rho, a_1,\pi$  etc poles, then according to
(151) in the correlators $\Pi^{V,A}(q,T)$  the poles do not shift in order
$T^2$  and appear at the same position as at $T=0$. The consequence of (151)
is  that, at $T\not= 0$  in transverse vector channel apart from the poles
corresponding to vector particle, there arise poles, corresponding to axial
particles and vice verca. In the same way a pion pole appears in the
longitudinal part of vector channel. The same phenomenon of parity mixing
(and, in some cases also isospin mixing)  appears at finite $T$  also in
other channels, including baryonic channels [63].

\section{Conclusion}

The goal of this review is to convince the reader, that chiral effective
theory (CET)  of strong interactions is: on one hand a direct consequence of
QCD, of the chiral symmetry of QCD and its spontaneous violation; and on the
other hand, a very effective tool with high predictive power for solving the
problems of strong interactions at low energies. It was demonstrated, that
in QCD the masses of light quarks ($u,d$ and, in some extent, also $s$) are
small and in a good approximation, when these masses are neglected,  QCD is
chirally symmetric. However, the physical spectrum of real world (including
the vacuum state) does not possesses this symmetry: there is nonvanishing
(in the limit $m_u,m_d\to 0$)  symmetry violating quark condensate, and the
baryon masses are by no means small, in contradiction with chiral symmetry.
It was shown, that these two facts -- the large baryon masses and the
appearance of quark condensate are tightly interconnected: the first can be
expressed through the second. The violation of chiral symmetry on the
physical spectrum means that chiral symmetry is broken spontaneously. The
direct consequence of this fact is the appearance of massless Goldstone
bosons in the spectrum (pion in case of $SU(2)$  symmetry, where $u$  and
$d$-quarks are considered as massless and $s$-quark as massive). The known
symmetry of the theory and the existence of massless Goldstone bosons allows
one to construct CET, valid at low energies. CET is an effective theory,
what means, that, when going to the next approximation -- higher powers of
particle momenta -- new additional terms in the theory Lagrangian appear.

In the review CET Lagrangian in the first and second orders in momenta was
explicitly  constructed and its main features were discussed. On few
examples it was demonstrated that CET is very powerful in consideration of
low energy interactions of pions. Low energy sum rules, which are the
subject of direct experimental test, were presented. It was demonstrated,
that CET is very suitable tool for the study of QCD at finite temperature.
The indications for phase transitions in QCD were obtained from this study.

 I am very thankful to H.Leutwyler for enlightening discussion of various
 aspects of CET, I learned a lot from his papers and reviews on this
 subject. I am also very indebted to him for his hospitality at Bern.

 This work was made possible in part by Award  No. RP2-2247 of U.S. Civilian
 Research and Development Foundation for Independent States of Former
 Soviet Union (CRDF), by Russian Found of Basic Research grant
 00-02-17808, and INTAS Call 2000 Grant (Project 587).

\end{document}